\documentclass[twocolumn, amsmath,amssymb, nobibnotes, aps, prx, superscriptaddress, citeautoscript]{revtex4-2}

\usepackage[colorlinks=true,linkcolor=blue,citecolor=blue,urlcolor=blue]{hyperref}
\usepackage[T1]{fontenc}
\DeclareUnicodeCharacter{00A0}{ }	
\usepackage{graphicx}
\usepackage{gensymb}
\usepackage{amsmath,bm}
\usepackage{braket}
\usepackage[normalem]{ulem}
\usepackage[shortlabels]{enumitem}
\usepackage[labelfont=bf]{caption}
\usepackage{titlesec}
\usepackage{setspace}

\usepackage{dcolumn}
\usepackage{subfigure}
\usepackage{natbib} 
\usepackage{multirow}
\usepackage{soul}
\usepackage{float}
\usepackage[normalem]{ulem}
\usepackage[usenames,dvipsnames]{color}
\usepackage{appendix}
\usepackage{mathrsfs}
\usepackage[colorlinks=true,linkcolor=blue,citecolor=blue,urlcolor=blue]{hyperref}
\usepackage{mathrsfs}

\begin{document}

\title{Accurate electronic properties and intercalation voltages of olivine-type Li-ion cathode materials from extended Hubbard functionals}

\author{Iurii Timrov}\email[]{ iurii.timrov@epfl.ch}
\affiliation{Theory and Simulation of Materials (THEOS), and National Centre for Computational Design and Discovery of Novel Materials (MARVEL), \'Ecole Polytechnique F\'ed\'erale de Lausanne (EPFL), CH-1015 Lausanne, Switzerland}

\author{Francesco Aquilante}
\affiliation{Theory and Simulation of Materials (THEOS), and National Centre for Computational Design and Discovery of Novel Materials (MARVEL), \'Ecole Polytechnique F\'ed\'erale de Lausanne (EPFL), CH-1015 Lausanne, Switzerland}

\author{Matteo Cococcioni}
\affiliation{Department of Physics, University of Pavia, via Bassi 6, I-27100 Pavia, Italy}

\author{Nicola Marzari}
\affiliation{Theory and Simulation of Materials (THEOS), and National Centre for Computational Design and Discovery of Novel Materials (MARVEL), \'Ecole Polytechnique F\'ed\'erale de Lausanne (EPFL), CH-1015 Lausanne, Switzerland}
\affiliation{Laboratory for Materials Simulations, Paul Scherrer Institut, 5232 Villigen PSI, Switzerland}

\begin{abstract} 
The design of novel cathode materials for Li-ion batteries would greatly benefit from accurate first-principles predictions of structural, electronic, and magnetic properties as well as intercalation voltages in compounds containing transition-metal elements. For such systems, density-functional theory (DFT) with standard (semi-)local exchange-correlation functionals is of limited use as it often fails due to strong self-interaction errors that are especially relevant in the partially filled $d$ shells. Here, we perform a detailed comparative study of the phospho-olivine cathode materials Li$_x$MnPO$_4$, Li$_x$FePO$_4$, and the mixed transition metal Li$_x$Mn$_{1/2}$Fe$_{1/2}$PO$_4$ ($x=0, 1/4, 1/2, 3/4, 1$) using four electronic-structure methods: DFT, DFT+$U$, DFT+$U$+$V$, and HSE06. We show that DFT+$U$+$V$, with onsite $U$ and intersite $V$ Hubbard parameters determined from first principles and self-consistently with respect to the structural parameters by means of density-functional perturbation theory (linear response), provides the most accurate description of the electronic structure of these challenging compounds. In particular, we demonstrate that DFT+$U$+$V$ displays very clearly ``digital'' changes in oxidation states of the transition-metal ions in all compounds, including the mixed-valence phases occurring at intermediate Li concentrations, leading to voltages in remarkable agreement with experiments. We show that the inclusion of intersite Hubbard interactions is essential for the accurate prediction of thermodynamic quantities, balancing the drive for localization induced by the onsite $U$ with intersite $V$ orbital hybridizations. At variance with other methods, DFT+$U$+$V$ describes accurately such localization-hybridization interplay, and thus opens the door for the study of more complex cathode materials as well as for a reliable exploration of the chemical space of compounds for Li-ion batteries.
\end{abstract}

\date{\today} 

\maketitle

\section{INTRODUCTION}
\label{sec:intro}

Recent years have witnessed urgent needs for renewable energy and the availability of energy storage technology that is needed at all scales. One of the major advances in this area can be traced back to the development of Li-ion rechargeable batteries~\cite{Goodenough:2013, Tarascon:2018} that are currently employed in a variety of applications, e.g. for portable electronics, power tools, automotive industry, electricity grids, to name a few~\cite{Tarascon:2014, Kang:2009}. These technologies are in increasing demand due to a global increase in energy consumption, widening dependence on the availability of efficient, safe, and nontoxic Li-ion batteries. 
 
The properties and performance of Li-ion batteries (such as power and energy density, capacity retention, cyclability, thermal stability, etc.) depend on many factors and their interplay within the complexity of the actual multicomponent devices. As part of this network, cathode materials play a pivotal role, determining the Li intercalation voltage and cyclability of Li$^+$ ions through the interface with the electrolyte. There are various types of cathode materials, among which we mention layered, spinel, olivine, prussian blue, and cation-disordered rocksalt~\cite{Monconduit:2015, Lee:2014, Hurlbutt:2018}. A key ingredient of cathodes are transition-metal (TM) elements that are electrochemically active species that change their oxidation state during charging and discharging of the battery. It is therefore of paramount importance to understand at the atomistic level what properties of such compounds lead to efficient electrochemical processes.   

An important tool for studying cathode materials is density-functional theory (DFT)~\cite{Hohenberg:1964,Kohn:1965}, which is a workhorse for first-principles simulations in physics, chemistry, and materials science. DFT requires approximations to the exchange-correlation (xc) functional, with local spin-density approximation (LSDA) and spin-polarized generalized-gradient approximation ($\sigma$-GGA) being the most popular ones. However, these approximations often provide some unsatisfactory results (e.g. voltages, formation energies, change in the atomic occupations in mixed-valence compounds, etc.) for many TM compounds due to self-interaction errors (SIE)~\cite{Perdew:1981, MoriSanchez:2006} which are especially large for localized $d$ and $f$ electrons. For this reason, more accurate approaches beyond ``standard DFT'' (i.e. based on LSDA or $\sigma$-GGA) are generally used, among which we mention Hubbard-corrected DFT based on LSDA or $\sigma$-GGA (so-called LSDA+$U$ and GGA+$U$~\cite{anisimov:1991, Liechtenstein:1995, Dudarev:1998} and its extensions LSDA+$U$+$V$ and GGA+$U$+$V$~\cite{Campo:2010, TancogneDejean:2020, Lee:2020} - in the following we will refer to these broadly as DFT+$U$ and DFT+$U$+$V$), meta-GGA functionals such as SCAN~\cite{Sun:2015} (and its flavors~\cite{Bartok:2019, Furness:2020}) and SCAN+$U$~\cite{Gautam:2018, Long:2020, Kaczkowski:2021, Artrith:2022}, DFT with hybrid functionals (e.g. PBE0~\cite{Adamo:1999} and HSE06~\cite{Heyd:2003, Heyd:2006}), to name a few. In DFT+$U$, the Hubbard $U$ correction is applied selectively only to the partially filled $d$ states of TM elements to alleviate SIE for these states~\cite{Kulik:2008}, while all other states are treated at the level of LSDA or $\sigma$-GGA.  In contrast, in meta-GGA functionals the kinetic energy density is taken into account and known exact constraints are satisfied (17 in the case of SCAN). Finally, in hybrid functionals a fraction of Fock exchange is added (25$\%$ in the case of PBE0 and HSE06) and the remainder of exchange is treated at the $\sigma$-GGA level, together with 100$\%$ of the $\sigma$-GGA correlation. In the context of a first-principle prediction of the properties of cathode materials, it still remains to establish which of these classes of advanced functionals provides the most accurate, reliable, and computationally affordable results.

The major interest in the use of Hubbard-corrected DFT comes from its ability to greatly improve the accuracy of  standard DFT with only a marginal increment in the computational cost~\cite{Himmetoglu:2014}. However, this is true only if the proper values of the Hubbard parameters are employed. In practice, these are unknown {\em a priori} and need to be determined by means of a robust protocol. One strategy that is widespread is to assign {\em bona fide} empirical values to the Hubbard parameters. For example, DFT+$U$ with $U$ parameters fitted to experimental binary metal formation energies~\cite{Wang:2006} using the Kubaschewski tables~\cite{Kubaschewski:1993} have proven to be effective for high-throughput search of novel cathode materials~\cite{Hautier:2011, Hautier:2011b, Mueller:2011, Ong:2011b}. Also, $U$ parameters are often calibrated empirically so that DFT+$U$ calculations reproduce some properties of interest (e.g. band gaps, magnetic moments, lattice parameters, oxidation enthalpies - see e.g. Refs.~\cite{LeBacq:2004, Aykol:2014, Isaacs:2017}), and are used to predict other properties, (e.g. voltages, formation energies, migration barriers, etc.). If, on one side, experimental results might not be available for the Hubbard parameters to be fitted on, on the other the so-tuned $U$ values are not always guarantied to be suitable for accurate predictions of other properties. Hence, finding empirically a global $U$ parameter that makes accurate predictions on many properties of a given material at the same time is a nontrivial task. In this respect, an alternative and very attractive approach is to compute these parameters using first-principle methods~\cite{Hubparam:2022}, such as constrained DFT (cDFT)~\cite{Dederichs:1984, Mcmahan:1988, Gunnarsson:1989, Hybertsen:1989, Gunnarsson:1990, Pickett:1998, Solovyev:2005, Nakamura:2006, Shishkin:2016}, Hartree-Fock based approaches~\cite{Mosey:2007, Mosey:2008, Andriotis:2010, Agapito:2015, TancogneDejean:2020, Lee:2020}, and the constrained random phase approximation (cRPA)~\cite{Springer:1998, Kotani:2000, Aryasetiawan:2004, Aryasetiawan:2006}. A linear-response (LR) formulation~\cite{Cococcioni:2005} of cDFT (LR-cDFT) has become a method of choice for many computational Hubbard-corrected DFT studies~\cite{Cococcioni:2019, Moore:2022}; moreover, its recent reformulation in terms of density-functional perturbation theory (DFPT)~\cite{Timrov:2018, Timrov:2021} further boosted its success thanks to the fact that it allows to replace computationally expensive supercells by a primitive cell with monochromatic perturbations, thus significantly reducing the computational burden of determining the Hubbard parameters. DFT+$U$ with Hubbard $U$ computed using LR-cDFT~\cite{Zhou:2004, Zhou:2004b, Zhou:2004c, Shishkin:2016} or cRPA~\cite{Kim:2021} has proven to be effective in improving intercalation voltages and electronic structure properties of cathode materials, and, remarkably, DFT+$U$+$V$ with $U$ and $V$ determined from LR-cDFT in a self-consistent fashion~\cite{Timrov:2021} was shown to provide excellent agreement with experimental voltages for olivine-type cathode materials~\cite{Cococcioni:2019} thereby highlighting the importance of intersite Hubbard $V$ interactions. Finally, Hubbard-corrected DFT calculations are sometimes performed including van der Waals (vdW) corrections (especially for layered materials) that were shown to further improve the accuracy of the computed properties~\cite{Aykol:2015}.

The meta-GGA SCAN functional has gained a lot of interest since its introduction in 2015~\cite{Sun:2015}, in particular for modeling of cathode materials. However, it is important to note that despite being very successful for a broad class of materials and properties, SCAN still contains significant SIE especially when applied to TM compounds~\cite{Kitchaev:2016, Hinuma:2017}, it exhibits some potential limitations in describing magnetic systems~\cite{Ekholm:2018, Tran:2020}, and it suffers from strong numerical instabilities~\cite{Lehtola:2022}. Moreover, from a technical point of view there are currently only few SCAN-based pseudopotentials (PPs)~\cite{SCAN_PPs} and often GGA-based PPs are used; this inconsistency is known to introduce noticeable errors in calculating some properties of materials as e.g. phase transition energies~\cite{Yao:2017}. Nonetheless, recent applications of SCAN have shown that it gives improved description of electronic properties and voltages in layered cathode materials compared to other functionals~\cite{Chakraborty:2018}; however, SCAN does not eliminate the need of the Hubbard $U$ correction in olivine and spinel materials~\cite{Isaacs:2020}. In fact, SCAN with the Hubbard $U$ correction (SCAN+$U$) was shown to give more accurate predictions for many transition-metal compounds than SCAN~\cite{Gautam:2018, Long:2020, Kaczkowski:2021, Artrith:2022}. In addition, SCAN and SCAN+$U$ are also used including the vdW corrections~\cite{Zhao:2022, Peng:2017, Isaacs:2020, Long:2021}; at present, only SCAN+\textit{r}VV10+$U$ is used since the revised Vydrov-Van Voorhis (\textit{r}VV10) functional~\cite{Vydrov:2010, Sabatini:2013} is the only vdW functional that has been parametrized for SCAN so far~\cite{Peng:2016}.

Finally, hybrid functionals exhibit a similar accuracy improvement over standard DFT as they work in the same direction of reducing SIE for TM compounds~\cite{Chevrier:2010}. At variance with Hubbard functionals and SCAN, however, their use comes at a much higher computational cost than standard DFT. Furthermore, for hybrid functionals, quite often the required fraction of Fock exchange must be tuned in solids in order to reach improved agreement with experiments~\cite{Seo:2015}. This approach is no less arbitrary than picking a $U$ value empirically. Although there are ways to determine the optimal amount of Fock exchange \textit{ab initio} needed for each system of interest~\cite{Skone:2014, Skone:2016, Bischoff:2019, Kronik:2012, Wing:2021, Lorke:2020}, very often the use of tuned hybrid functional tends to deviate considerably from a pure first-principle based practice. 
If one disregards the option of tuning the amount of Fock exchange, it remains the possibility to choose the hybrid functional upon considerations of its reliability for the problem of interest. In particular, for the study of cathode materials HSE06 has proven its ability to predict accurate electronic and electrochemical properties of some paradigmatic examples of such systems as, e.g. the phospho-olivine Li$_x$MnPO$_4$~\cite{Ong:2011} and the spinel Li$_x$Mn$_2$O$_4$~\cite{Eckhoff:2020}. The possibility to use hybrids with vdW corrections guarantees their applicability for the study of layered and organic systems~\cite{Peelaers:2014, Zhang:2017b}.

In this paper, we present a detailed comparative study for selected olivine-type cathode materials: Li$_x$FePO$_4$, Li$_x$MnPO$_4$, as well as the more complex mixed-TM compound Li$_x$Mn$_{1/2}$Fe$_{1/2}$PO$_4$ ($x=0, 1/4, 1/2, 3/4, 1$). We perform calculations with the four electronic-structure methods: DFT, DFT+$U$, DFT+$U$+$V$, and HSE06, with the aim to assess the reliability of their predictions (e.g., oxidation states and Li intercalation voltages) in comparison with experiments. We show that DFT+$U$+$V$ remarkably outperforms the other three well-established methods. A key requirement is that the onsite $U$ and intersite $V$ Hubbard parameters are determined from first-principles self-consistently using DFPT~\cite{Timrov:2018, Timrov:2021}. In particular, we demonstrate that DFT+$U$+$V$ accurately predicts the electronic structure not only for the fully delithiated/lithiated compounds ($x=0, 1$), but also for the intermediate Li concentrations ($x=1/4, 1/2, 3/4$). Overall, HSE06 and DFT+$U$ results are in good qualitative agreement with 
DFT+$U$+$V$, and superior to standard DFT, but intercalation voltages do not match the quantitative accuracy shown by DFT+$U$+$V$. 
Importantly, DFT+$U$+$V$ predicts a ``digital'' change in atomic occupations when gradually changing the Li concentration while DFT averages out the occupations over sites and HSE06 shows a less clear pattern in the ``digital'' change of occupations for Fe-containing phospho-olivines. This study shows that the inclusion of intersite interactions $V$ is essential for the accurate prediction of thermodynamic quantities when electronic localization occurs in the presence of significant interatomic hybridization, confirming and enriching the findings of earlier work by some of us~\cite{Cococcioni:2019}. 

The paper is organized as follows. Section~\ref{sec:methods} presents the theoretical framework with the basics of HSE06, DFT+$U$, and DFT+$U$+$V$, and the linear-response calculation of $U$ and $V$ using DFPT. In Sec.~\ref{sec:results_and_discussion} we present our findings for the oxidation states, L\"owdin occupations, spin-resolved projected density of states (PDOS), and voltages; and in Sec.~\ref{sec:conclusions} we provide the conclusions.

\section{METHODS}
\label{sec:methods}

In this section we discuss the basics of HSE06~\cite{Heyd:2003, Heyd:2006}, DFT+$U$~\cite{anisimov:1991, Dudarev:1998}, and DFT+$U$+$V$~\cite{Campo:2010, Himmetoglu:2014} as well as the main idea of the DFPT approach for computing Hubbard parameters~\cite{Timrov:2018, Timrov:2021}. In the following, we use the generic name ``DFT+Hubbard'' as broadly referring to any flavor of Hubbard-corrected DFT, which in this paper covers DFT+$U$ and DFT+$U$+$V$. For the sake of simplicity, the formalism is presented in the framework of norm-conserving (NC) PPs in the collinear spin-polarized case. Hartree atomic units are used throughout.

\subsection{DFT+Hubbard}

For the sake of generality, here we discuss the DFT+$U$+$V$ formalism~\cite{Campo:2010}. It can be easily simplified to DFT+$U$ by setting $V=0$. In DFT+$U$+$V$, the xc energy contains a (semi-)local functional (e.g., PBEsol~\cite{Perdew:2008}) and a corrective Hubbard term~\cite{Campo:2010}: 
\begin{equation}
E_\mathrm{xc}^{\mathrm{PBEsol}+U+V} = E_\mathrm{xc}^{\mathrm{PBEsol}} + E_\mathrm{xc}^{U+V} \,,
\label{eq:Edft_plus_u}
\end{equation}
where $E_\mathrm{xc}^{U+V}$ is the Hubbard energy that removes (from the Hubbard manifold) SIE present due to the use of approximations in the xc functional. At variance with the DFT+$U$ approach whose Hubbard corrective term only contains onsite interactions (scaled by $U$), DFT+$U$+$V$ also features intersite interactions (scaled by $V$) between orbitals centered on different sites. In the simplified rotationally-invariant formulation, the extended Hubbard term $E_\mathrm{xc}^{U+V}$ is defined such that it removes the mean-field PBEsol-based interactions in the Hubbard manifold and adds the ones that restore the piecewise linear energy behavior~\cite{Himmetoglu:2014}, and it reads~\cite{Campo:2010}:
\begin{eqnarray}
E_\mathrm{xc}^{U+V} & = & \frac{1}{2} \sum_I \sum_{\sigma m m'} 
U^I \left( \delta_{m m'} - n^{II \sigma}_{m m'} \right) n^{II \sigma}_{m' m} \nonumber \\
& & - \frac{1}{2} \sum_{I} \sum_{J (J \ne I)}^* \sum_{\sigma m m'} V^{I J} 
n^{I J \sigma}_{m m'} n^{J I \sigma}_{m' m} \,,
\label{eq:Edftu}
\end{eqnarray}
where $\sigma$ is the spin index, $I$ and $J$ are atomic site indices, $m$ and $m'$ are the magnetic quantum numbers associated with a specific angular momentum, $U^I$ and $V^{I J}$ are the effective onsite and intersite Hubbard parameters, and the asterisk in the sum denotes that for each atom $I$, the index $J$ covers all its neighbors up to a given distance (or up to a given shell). As apparent from its expression, by subtracting a term quadratic in the atomic occupations and substituting it with a linear one, the Hubbard correction contributes to decreasing the curvature of the energy as a function of the occupations of the Hubbard manifold (a measure of the effective self-interaction) and to reestablish a piecewise linear behavior~\cite{Himmetoglu:2014}. While such piecewise linearity is not a formal requirement of energy functionals, it has been long argued~\cite{Cococcioni:2005, Kulik:2006, Kulik:2008, Kulik:2011} that it is an essential condition to reduce SIE in systems with very localized (e.g., $d$ and $f$) electrons. The generalized occupation matrices $n^{I J \sigma}_{m m'}$ are based on a projection of the Kohn-Sham (KS) wave functions $\psi^\sigma_{v,\mathbf{k}}(\mathbf{r})$ on localized orbitals $\varphi^{I}_{m}(\mathbf{r})$ of neighbor atoms~\cite{Campo:2010}: 
\begin{equation}
n^{I J \sigma}_{m m'} = \sum_{v,\mathbf{k}} f^\sigma_{v,\mathbf{k}}
\langle \psi^\sigma_{v,\mathbf{k}} | \varphi^{J}_{m'} \rangle \langle \varphi^{I}_{m} | \psi^\sigma_{v,\mathbf{k}} \rangle \,, 
\label{eq:occ_matrix_0}
\end{equation}
where $v$ is the electronic band index, $\mathbf{k}$ indicates points in the first Brillouin zone, $f^\sigma_{v,\mathbf{k}}$ are the occupations of the KS states, and $\varphi^I_{m}(\mathbf{r}) \equiv \varphi^{\gamma(I)}_{m}(\mathbf{r} - \mathbf{R}_I)$ are localized orbitals centered on the $I$th atom of type $\gamma(I)$ at the position $\mathbf{R}_I$. It is convenient to establish a shorthand notation for the onsite occupation matrix: $n^{I\sigma}_{m m'} \equiv n^{II\sigma}_{m m'}$, which is used in DFT+$U$ corresponding to the first line of Eq.~\eqref{eq:Edftu}. The two terms in Eq.~\eqref{eq:Edftu} (i.e., proportional to the onsite $U^{I}$ and intersite $V^{IJ}$ couplings) counteract each other: the onsite term favors localization on atomic sites (thus suppressing hybridization with neighbors), while the intersite term favors hybridized states with components on neighbor atoms (thus suppressing the onsite localization). It is the balance between these two competing effects that determines the ground state of the system. Therefore, an accurate evaluation of $U^{I}$ and $V^{IJ}$ is crucial in this respect.

In DFT+Hubbard the values of the Hubbard parameters are not known {\it a~priori}, and hence they are often adjusted empirically such that the final results of simulations match some experimental properties of interest (e.g. band gaps, oxidation enthalpies, etc.). This procedure introduces a degree of arbitrariness (e.g. on the choice of experimental measurements to match) and indeterminacy (there might be several sets of interaction parameters able to reproduce a limited number of experimental results) and makes DFT+Hubbard not fully first principles. Sometimes the match to experimental measurements is also questionable on conceptual grounds, e.g. when a band gap is matched, given that exact DFT would also not reproduce the experimental gap. Most importantly, it restricts the applicability of this corrective scheme only to a domain of materials for which the Hubbard parameters can be validated with experimental results and limits its use for investigating the behavior of not-yet synthesized systems. 
Moreover, it is often forgotten that the Hubbard $U$ correction is applied using Hubbard projectors that can be defined in many different ways~\cite{Timrov:2020b}, e.g. taken from the atomic calculations used to generate the respective pseudopotentials or their orthogonalized counterparts (see Eq.~\eqref{eq:OAO_def}), and that can be constructed with different degrees of oxidation. Hence, these Hubbard projectors and the respective $U$ parameters are not transferable and one should not consider $U$ as a universal number for a given element or material (see the appendix in Ref.~\cite{Kulik:2008}). Therefore, a first-principles calculation of the Hubbard parameters is essential for quantitative reliability and thus highly desirable.

In many cases, where localization occurs on atomic states, the effect of a finite $V$ might actually be mimicked by a smaller value of $U$ that avoids suppressing intersite hybridization too much. However, there are cases where localization might actually occur on bonds~\cite{Campo:2010}, and in these cases the use of an intersite $V$ cannot be mimicked by any small value of $U$ which lacks the physics needed for intersite covalent bonding. Therefore, DFT+$U$+$V$ where both $U$ and $V$ values are computed from first principles constitutes a robust and accurate approach that describes accurately the onsite localization and intersite hybridization of electrons without any manual calibrations of Hubbard parameters.

The aforementioned LR-cDFT approach allows us to compute $U$ and $V$ from a generalized piece-wise linearity condition~\cite{Cococcioni:2005, Campo:2010}. Within this framework the Hubbard parameters are the elements of an effective interaction matrix computed as the difference between bare and screened inverse susceptibilities~\cite{Cococcioni:2005}:
\begin{equation}
U^I = \left(\chi_0^{-1} - \chi^{-1}\right)_{II} \,,
\label{eq:Ucalc}
\end{equation}
\begin{equation}
V^{IJ} = \left(\chi_0^{-1} - \chi^{-1}\right)_{IJ} \,.
\label{eq:Vcalc}
\end{equation}
The susceptibility matrices $\chi_0$ and $\chi$ measure the response of atomic occupations to a shift in the potential acting on the atomic states of a specific Hubbard atom~\cite{Cococcioni:2005}: $\chi_{IJ} = \sum_{m\sigma} \left(dn^{I \sigma}_{mm} / d\alpha^J\right)$. The difference between $\chi_0$ and $\chi$ consists in the fact that the former represents the (bare) response to the total potential (i.e., before the electronic charge density readjusts self-consistently), while the latter -- the (total) response to the external potential~\cite{Linscott:2018}. In order to avoid  computationally demanding supercell calculations, required within the LR-cDFT approach to make the perturbation isolated, we have recently recast the LR calculation outlined above within DFPT, so that the response to isolated perturbations can be efficiently computed from the superposition of the variation of atomic occupations to monochromatic (i.e., wave-vector-specific) perturbations using primitive cells~\cite{Timrov:2018}:
\begin{equation}
\frac{dn^{I \sigma}_{mm'}}{d\alpha^J} = \frac{1}{N_{\mathbf{q}}}\sum_{\mathbf{q}}^{N_{\mathbf{q}}} e^{i\mathbf{q}\cdot(\mathbf{R}_{l} - \mathbf{R}_{l'})}\Delta_{\mathbf{q}}^{s'} n^{s \sigma}_{mm'} \,.
\label{eq:dnq}
\end{equation}
Here, $\Delta_{\mathbf{q}}^{s'} n^{s \sigma}_{mm'}$ is the response of the occupation matrix, $I \equiv (l, s)$ and $J \equiv (l', s')$, where $s$ and $s'$ are the atomic indices in unit cells while $l$ and $l'$ are the unit-cell indices. $\mathbf{R}_l$ and $\mathbf{R}_{l'}$ are the Bravais lattice vectors, and the grid of $\mathbf{q}$ points is chosen fine enough to make the resulting atomic perturbations effectively decoupled from their periodic replicas. An exhaustive illustration of this approach can be found in Refs.~\cite{Timrov:2018, Timrov:2021}, where a recent extension to ultrasoft pseudopotentials and to the projector-augmented-wave method is also discussed. 

It is crucial to keep in mind that the values of the computed Hubbard parameters strongly dependent on the type of Hubbard projector functions $\varphi^I_m(\mathbf{r})$ that are used in DFT+Hubbard~\cite{Tablero:2008, Kulik:2008, Wang:2016, Timrov:2020b}. Here we employ orthogonalized atomic orbitals that are computed using the L\"owdin orthogonalization method~\cite{Lowdin:1950, Mayer:2002}:
\begin{equation}
    \varphi^I_{m}(\mathbf{r}) = \sum_{J m'} \left(\hat{O}^{-\frac{1}{2}}\right)^{JI}_{m' m} \phi^J_{m'}(\mathbf{r}) \,,
    \label{eq:OAO_def}
\end{equation}
where $\hat{O}$ is the orbital overlap matrix, whose matrix elements are defined as: $(\hat{O})^{IJ}_{m m'} = \langle \phi^I_{m} | \phi^J_{m'} \rangle$, and $\phi^I_{m}(\mathbf{r})$ are the nonorthogonalized atomic orbitals provided with PPs. With this choice of projector functions the electrons in the intersite overlap regions are not counted twice when computing the atomic occupations used in the Hubbard correction, as it is instead the case for the nonorthogonalized atomic orbitals $\phi^I_{m}(\mathbf{r})$. As a matter of fact, DFT+Hubbard with the L\"owdin orthogonalized orbitals has proven to give more accurate results for various properties of materials~\cite{Ricca:2020, Timrov:2020c, KirchnerHall:2021, Xiong:2021, Cococcioni:2021, Mahajan:2021, Mahajan:2022}, provided the Hubbard parameters are consistently computed with the L\"owdin orthogonalized orbitals. Therefore, Hubbard parameters and Hubbard projectors should always be defined consistently and reported together.

\subsection{HSE06}

In the range-separated hybrid functional HSE06 the exchange energy is divided into a short-range (S) and a long-range (L) part. Only 25$\%$ of the short-range part consists of the Fock energy and the remaining 75$\%$ is the PBE exchange energy, while the long-range exchange part is fully computed at the PBE level~\cite{Heyd:2003, Heyd:2006}. The total xc energy is
\begin{equation}
    E_\mathrm{xc}^\mathrm{HSE06} = \frac{1}{4} E_\mathrm{x}^\mathrm{Fock, S} + \frac{3}{4} E_\mathrm{x}^\mathrm{PBE, S} + E_\mathrm{x}^\mathrm{PBE, L} + E_\mathrm{c}^\mathrm{PBE} \,,
\end{equation}
where $E_\mathrm{c}^\mathrm{PBE}$ is the PBE correlation energy. The Fock short-range energy is the most computationally expensive term and it is defined by generalizing the definition of Fock~\cite{Kotani:1995} as follows:
\begin{eqnarray}
    E_\mathrm{x}^\mathrm{Fock, S} & = & -\frac{1}{2} \sum_{\sigma} \sum_{v,\mathbf{k}} \sum_{v',\mathbf{k}'} \int \int d\mathbf{r} d\mathbf{r}' \mathrm{erfc}(\omega |\mathbf{r}-\mathbf{r}'|) \nonumber \\
    & & \times \frac{\psi^{\sigma *}_{v,\mathbf{k}}(\mathbf{r}) \psi^{\sigma}_{v',\mathbf{k}'}(\mathbf{r}) \psi^{\sigma *}_{v',\mathbf{k}'}(\mathbf{r}') \psi^{\sigma}_{v,\mathbf{k}}(\mathbf{r}')}{|\mathbf{r}-\mathbf{r}'|} \,,
    \label{eq:EXX_def}
\end{eqnarray}
where $\mathrm{erfc}$ is the complementary error function, $\omega = 0.106 \, a_0^{-1}$ is the screening parameter with $a_0$ being the Bohr radius~\cite{Heyd:2006}. The Fock short-range energy of Eq.~\eqref{eq:EXX_def} contains only exchange interactions at relatively short atomic length scales, and it can be assimilated to a sort of ``onsite exchange'' and ``intersite exchange'' energy contributions: the former refers to exchange acting between orbitals centered on the same atom while the latter refers to exchange acting between orbitals centered on different atoms~\cite{Eckhoff:2020}. This aspect is crucial and in the following we further investigate such an analogy between HSE06 and DFT+Hubbard. 

\subsection{HSE06 versus DFT+Hubbard}

It is instructive to establish analogies between the hybrid functional HSE06 and DFT+Hubbard~\cite{Hubbard_vs_GW_vs_DFTU}. It has been shown in Refs.~\cite{TancogneDejean:2020, Lee:2020} that DFT+$U$+$V$ predicts the electronic structure of TM compounds and light-element compounds in closer agreement with HSE06 with respect to DFT+$U$. However, the origin of this improvement was not investigated in detail. As will be shown in what follows, DFT+$U$+$V$ is as accurate as (and occasionally better than) HSE06 in predicting electronic properties and voltages in phospho-olivines, and we provide a simple qualitative explanation for this.

DFT+$U$ and hybrid functionals share one important feature: they both attempt to correct SIE for orbitals centered on the same site. However, in contrast to HSE06, DFT+$U$ does not correct for SIE originating from the interactions of orbitals centered on different (neighboring) sites. This is why for systems with strong covalent bonding DFT+$U$ typically disagrees with HSE06 predictions. In contrast, this latter effect is captured by DFT+$U$+$V$, which makes it more general and allows us to cover similar physics as the one described by HSE06. However, it is important to recall that in DFT+$U$+$V$ only a subset of orbitals are corrected while hybrid functionals act on all the orbitals. Moreover, since typically only nearest-neighbor intersite interactions are taken into account in DFT+$U$+$V$, this looks similar to HSE06 that has only the short-range Fock exchange, while long-range effects are fully disregarded both in DFT+$U$+$V$ and HSE06.

\subsection{Crystal structure, magnetic ordering, and further details of calculation}
\label{sec:structure}

\begin{figure}[t]
  \includegraphics[width=0.9\linewidth]{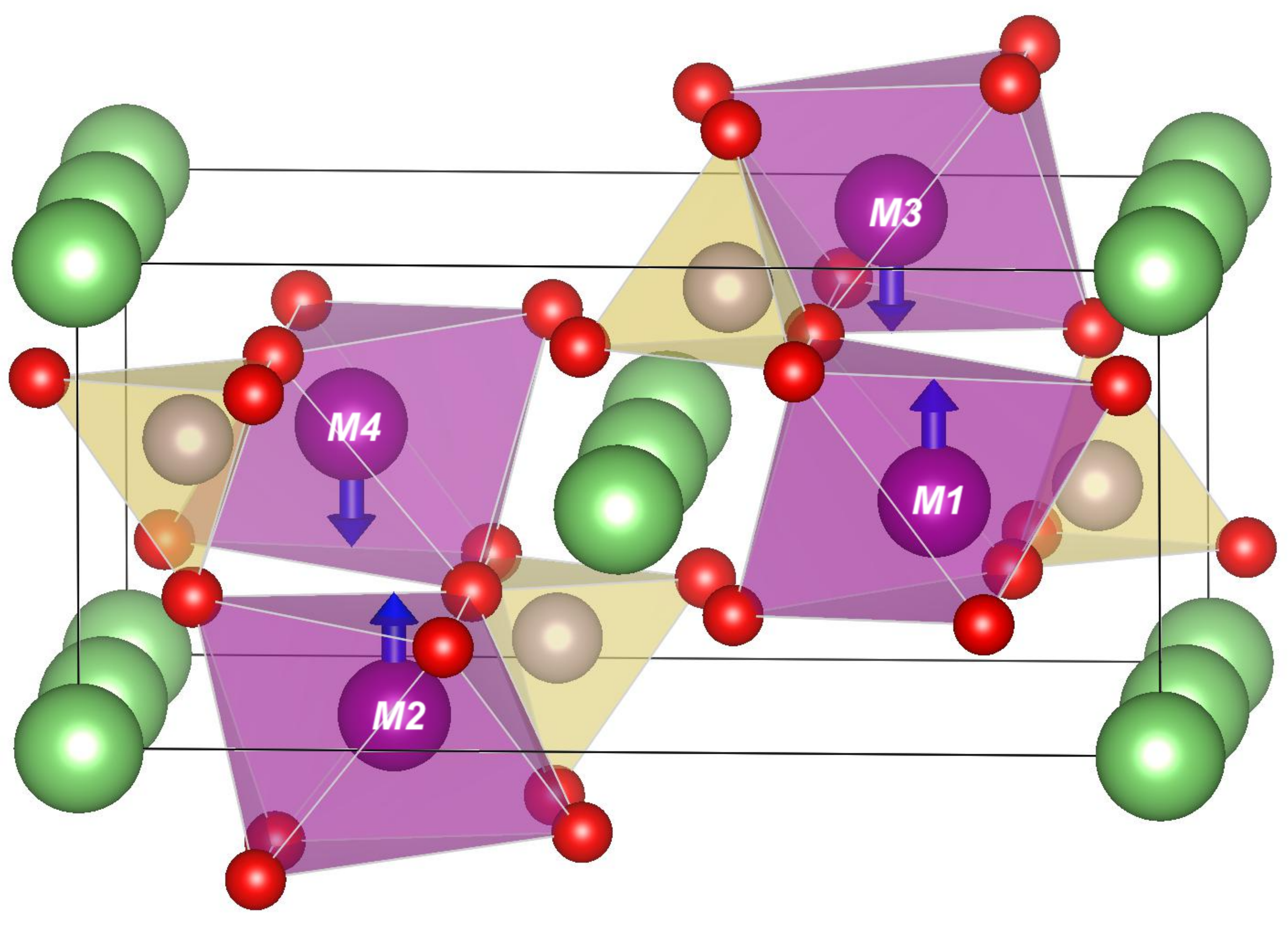}
   \caption{Crystal structure of phospho-olivines. Transition-metal elements ($M1$--$M4$) are indicated in purple, O atoms in red, Li atoms in green, and P atoms in yellow. Blue vertical arrows indicate the orientation of spin. In Li$_x$MnPO$_4$, $M1$--$M4$ correspond to Mn1--Mn4, in Li$_x$FePO$_4$, $M1$--$M4$ correspond to Fe1--Fe4, and in Li$_x$Mn$_{1/2}$Fe$_{1/2}$PO$_4$: $M1$=Fe1, $M2$=Mn2, $M3$=Fe3, $M4$=Mn4. Rendered using \textsc{VESTA}~\cite{Momma:2008}.}
\label{fig:crystal_structure}
\end{figure}

The phospho-olivines Li$_x$FePO$_4$, Li$_x$MnPO$_4$, and Li$_x$Mn$_{1/2}$Fe$_{1/2}$PO$_4$ have an orthorhombic crystal structure at $x=0$ and $x=1$ with a $Pnma$ space group~\cite{Nie:2010, Padhi:1997, Muraliganth:2010}. The unit cell contains four formula units, i.e. 24 atoms for $x=0$ and 28 atoms for $x=1$. The crystal structure of these systems is shown in Fig.~\ref{fig:crystal_structure} for $x=1$. The TM atoms (labeled as $M$ with an index from 1 to 4 in Fig.~\ref{fig:crystal_structure}) are coordinated by six O atoms forming a $M$O$_6$ octahedron of which it occupies the center. The P atoms are instead at the center of PO$_4$ tetrahedra that they form with neighboring oxygens. The three-dimensional structure of the crystal can be understood as being based on a network of corner-sharing $M$O$_6$ octahedra further linked by ``interstitial'' PO$_4$ tetrahedra that act as structural reinforcer [avoiding excessive volume variations upon Li (de-)intercalation] and chemical stabilizers (useful to avoid oxygen escapes). Li ions reside within octahedral channels parallel to the intermediate-length side of the cell. 

The phospho-olivines are known to show an antiferromagnetic behavior below their transition temperatures~\cite{Newnham:1965, Santoro:1967, Rousse:2003, Gnewuch:2020}. In the previous study (Ref.~\cite{Cococcioni:2019}) it was shown that different antiferromagnetic arrangements of spins result in total energies that differ not more than by $\sim 20$~meV at the DFT+Hubbard level of theory (largely irrelevant for the calculation of voltages). In this paper we use the magnetic configuration that minimizes the total energy (labeled ``AF$_1$'' in Ref.~\cite{Cococcioni:2019}), and it is depicted in Fig.~\ref{fig:crystal_structure}. Moreover, we use the same spin arrangement in the mixed TM phospho-olivine Li$_x$Mn$_{1/2}$Fe$_{1/2}$PO$_4$. Finally, there are several configurations for arranging two Mn and two Fe atoms in the unit cell of Li$_x$Mn$_{1/2}$Fe$_{1/2}$PO$_4$. Our goal here is not to investigate all configurations but rather to choose one as a representative case for comparing results obtained using different approaches. To this end, we choose to arrange Mn and Fe atoms such that two Mn atoms are antiferromagnetically coupled to each other and same for Fe atoms, as shown in Fig.~\ref{fig:crystal_structure}.

All technical details of the calculations are described in Sec.~S1 of the Supplemental Material (SM)~\cite{Note:SM:2022}. Hubbard parameters are computed self-consistently using DFPT as described in Sec.~\ref{sec:methods}, and their values are listed in Sec.~S2 of SM~\cite{Note:SM:2022}. The crystal structure is optimized using DFT and DFT+Hubbard, and the results are reported in Sec.~S3 of SM~\cite{Note:SM:2022}; for HSE06 calculations we use the DFT+$U$+$V$ geometry since the structural optimization at the HSE06 level is computationally too expensive. The configurations for partially delithiated structures and the formation energies are discussed in Sec.~S4. Other properties reported in the SM will be mentioned in Sec.~\ref{sec:results_and_discussion}.

\section{RESULTS AND DISCUSSION}
\label{sec:results_and_discussion}

\subsection{Oxidation states, L\"owdin occupations, and magnetic moments}
\label{sec:results-occupations_and_OS}

\begin{table*}[t]
\centering
\begin{tabular}{c|l|ccccccccccccccc}
\hline\hline
Material & \parbox{1.5cm}{Method} & \parbox{1cm}{$x$} & $\lambda_1^{\uparrow}$ & $\lambda_2^{\uparrow}$ & $\lambda_3^{\uparrow}$ & $\lambda_4^{\uparrow}$ & $\lambda_5^{\uparrow}$ & \phantom{a} & $\lambda_1^{\downarrow}$ & $\lambda_2^{\downarrow}$ & $\lambda_3^{\downarrow}$ & $\lambda_4^{\downarrow}$ & $\lambda_5^{\downarrow}$ & \parbox{1.3cm}{$n$} & \parbox{1.3cm}{$m$ ($\mu_\mathrm{B}$)} & \parbox{1cm}{OS} \\
\hline\hline
\parbox[t]{7mm}{\multirow{10}{*}{\rotatebox[origin=c]{90}{Li$_x$MnPO$_4$}}} 
& \multirow{2}{*}{DFT}         & 0 & 0.42       & {\bf 0.98} & {\bf 0.99} & {\bf 0.99} & {\bf 0.99} & & 0.09 & 0.10 & 0.13 & 0.16 & 0.27  & 5.12  &  3.63 & +3 \\
&                              & 1 & {\bf 0.99} & {\bf 0.99} & {\bf 0.99} & {\bf 1.00} & {\bf 1.00} & & 0.03 & 0.04 & 0.05 & 0.10 & 0.11  & 5.28  &  4.62 & +2 \\ \cline{2-17}
& \multirow{2}{*}{DFT+$U$}     & 0 & 0.54       & {\bf 0.99} & {\bf 0.99} & {\bf 1.00} & {\bf 1.00} & & 0.04 & 0.05 & 0.06 & 0.08 & 0.19  & 4.95  &  4.10 & +3 \\
&                              & 1 & {\bf 0.99} & {\bf 0.99} & {\bf 1.00} & {\bf 1.00} & {\bf 1.00} & & 0.02 & 0.02 & 0.03 & 0.07 & 0.08  & 5.19  &  4.76 & +2 \\ \cline{2-17}
& \multirow{2}{*}{DFT+$U$+$V$} & 0 & 0.50       & {\bf 0.99} & {\bf 0.99} & {\bf 1.00} & {\bf 1.00} & & 0.05 & 0.06 & 0.08 & 0.09 & 0.22  & 4.98  &  3.97 & +3 \\
&                              & 1 & {\bf 0.99} & {\bf 0.99} & {\bf 1.00} & {\bf 1.00} & {\bf 1.00} & & 0.02 & 0.02 & 0.03 & 0.07 & 0.08  & 5.21  &  4.75 & +2 \\ \cline{2-17}  
& \multirow{2}{*}{HSE06}       & 0 & 0.40       & {\bf 0.99} & {\bf 0.99} & {\bf 0.99} & {\bf 0.99} & & 0.06 & 0.07 & 0.09 & 0.10 & 0.23  & 4.91  &  3.83 & +3 \\
&                              & 1 & {\bf 0.99} & {\bf 0.99} & {\bf 1.00} & {\bf 1.00} & {\bf 1.00} & & 0.02 & 0.02 & 0.03 & 0.07 & 0.08  & 5.21  &  4.75 & +2 \\ \cline{2-17}
& \multirow{2}{*}{Nominal}     & 0 & 0.00       & {\bf 1.00} & {\bf 1.00} & {\bf 1.00} & {\bf 1.00} & & 0.00 & 0.00 & 0.00 & 0.00 & 0.00  & 4.00  &  4.00 & +3 \\
&                              & 1 & {\bf 1.00} & {\bf 1.00} & {\bf 1.00} & {\bf 1.00} & {\bf 1.00} & & 0.00 & 0.00 & 0.00 & 0.00 & 0.00  & 5.00  &  5.00 & +2 \\
\hline
\parbox[t]{7mm}{\multirow{10}{*}{\rotatebox[origin=c]{90}{Li$_x$FePO$_4$}}} 
& \multirow{2}{*}{DFT}         & 0 & {\bf 0.97} & {\bf 0.98} & {\bf 0.99} & {\bf 1.00} & {\bf 1.00} & & 0.15 & 0.16 & 0.17 & 0.25 & 0.26        & 5.93  &  3.94 & +3 \\
&                              & 1 & {\bf 0.99} & {\bf 0.99} & {\bf 0.99} & {\bf 0.99} & {\bf 1.00} & & 0.06 & 0.07 & 0.13 & 0.14 & {\bf 0.98}  & 6.32  &  3.57 & +2 \\ \cline{2-17}
& \multirow{2}{*}{DFT+$U$}     & 0 & {\bf 0.99} & {\bf 0.99} & {\bf 1.00} & {\bf 1.00} & {\bf 1.00} & & 0.09 & 0.10 & 0.10 & 0.22 & 0.24        & 5.72  &  4.22 & +3 \\
&                              & 1 & {\bf 0.99} & {\bf 0.99} & {\bf 1.00} & {\bf 1.00} & {\bf 1.00} & & 0.02 & 0.04 & 0.08 & 0.09 & {\bf 1.00}  & 6.20  &  3.76 & +2 \\ \cline{2-17}
& \multirow{2}{*}{DFT+$U$+$V$} & 0 & {\bf 0.99} & {\bf 0.99} & {\bf 1.00} & {\bf 1.00} & {\bf 1.00} & & 0.09 & 0.12 & 0.12 & 0.21 & 0.25        & 5.76  &  4.18 & +3 \\
&                              & 1 & {\bf 0.99} & {\bf 0.99} & {\bf 1.00} & {\bf 1.00} & {\bf 1.00} & & 0.03 & 0.04 & 0.09 & 0.10 & {\bf 0.99}  & 6.22  &  3.74 & +2 \\ \cline{2-17}  
& \multirow{2}{*}{HSE06}       & 0 & {\bf 0.99} & {\bf 0.99} & {\bf 0.99} & {\bf 0.99} & {\bf 1.00} & & 0.09 & 0.10 & 0.10 & 0.19 & 0.23        & 5.67  &  4.26 & +3 \\
&                              & 1 & {\bf 0.99} & {\bf 0.99} & {\bf 1.00} & {\bf 1.00} & {\bf 1.00} & & 0.03 & 0.04 & 0.09 & 0.09 & {\bf 0.99}  & 6.22  &  3.74 & +2 \\ \cline{2-17}
& \multirow{2}{*}{Nominal}     & 0 & {\bf 1.00} & {\bf 1.00} & {\bf 1.00} & {\bf 1.00} & {\bf 1.00} & & 0.00 & 0.00 & 0.00 & 0.00 & 0.00        & 5.00  &  5.00 & +3 \\
&                              & 1 & {\bf 1.00} & {\bf 1.00} & {\bf 1.00} & {\bf 1.00} & {\bf 1.00} & & 0.00 & 0.00 & 0.00 & 0.00 & {\bf 1.00}  & 6.00  &  4.00 & +2 \\
\hline\hline
\end{tabular}%
\caption{Population analysis data for the $3d$ shells of Mn and Fe atoms in Li$_x$MnPO$_4$ and Li$_x$FePO$_4$ at $x=0$ and $x=1$ computed using four approaches: DFT (PBEsol functional), DFT+$U$, DFT+$U$+$V$, HSE06, and the nominal data. This table shows the eigenvalues of the site-diagonal occupation matrix for the spin-up ($\lambda_i^\uparrow$, $i=\overline{1,5}$) and spin-down ($\lambda_i^\downarrow$, $i=\overline{1,5}$) channels, L\"owdin occupations $n = \sum_i (\lambda_i^\uparrow + \lambda_i^\downarrow)$, magnetic moments $m = \sum_i (\lambda_i^\uparrow - \lambda_i^\downarrow)$, and the oxidation state (OS). For the sake of simplicity we dropped the atomic site index $I$ from all quantities reported here. The eigenvalues are written in the ascending order (from left to right) for each spin channel. The eigenvalues written in bold correspond to fully occupied states and thus are taken into account when determining the OS according to Ref.~\cite{Sit:2011}.}
\label{tab:OS_Mn_and_Fe}
\end{table*}

\begin{figure*}[t!]
  \includegraphics[width=0.85\linewidth]{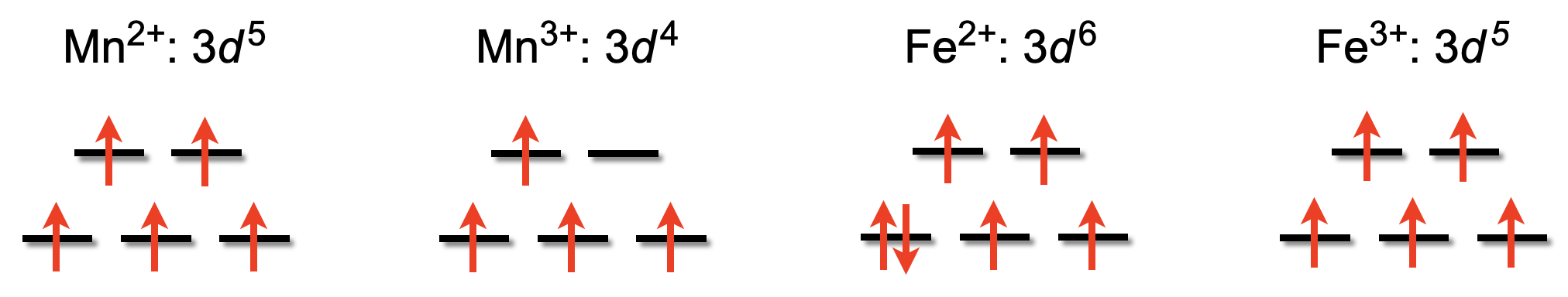}
   \caption{Nominal occupations of the $3d$ manifold of Mn and Fe atoms (not hybridized with ligands) in a high-spin undistorted octahedral complex  with different oxidation states ($O_h$ point group). The $t_{2g}$ and $e_g$ levels are indicated with black horizontal lines and are nondegenerate due to the crystal-field splitting; up and down red arrows correspond to spin-up and spin-down electrons, respectively.}
\label{fig:Nominal_OS}
\end{figure*}

The concept of oxidation state (OS) is central and ubiquitous in chemistry and physics, it is widely used to describe redox reactions, electrolysis, and many electrochemical processes as it allows to track electron movement during reactions~\cite{Jorgensen:1969}. The main idea is that the variations in electron number must be integer and this assigns the OS of an ion~\cite{Goldbook:2021}.
 
However, OS has long eluded a proper quantum-mechanical interpretation. Numerous methods have been proposed to determine the OS, and such methods often infer the OS of ions from schemes for allocating charges to ions. These schemes can be classified into categories, among which we mention: $i)$~partition of space with integration of the total charge density within space allocated to each ion (e.g. Bader~\cite{Bader:1990} and Voronoi~\cite{Bickelhaupt:1996} charges), and $ii)$~projection of the electronic wavefunctions onto a localized basis (e.g. Mulliken~\cite{Mulliken:1955} and L\"owdin~\cite{Lowdin:1950} charges, or natural bond orbitals~\cite{Reed:1988}). On the one hand, in the partition schemes all orbitals contribute to the charge within the allocated volume (e.g. a sphere of a certain radius centered on an ion), thus losing the connection to the OS of individual ions and its certain manifold (e.g. $d$ orbitals of TM elements). On the other hand, projection schemes present a dependence on the type of projector functions that are used to construct the localized basis set (and dependence on the cutoff radii used in some methods). The electronic populations computed using these methods are quite useful to give an indication of the OS, however these populations are often noninteger and their changes during redox reactions are significantly smaller than the changes in the nominal electron numbers for the end elements of the reaction. Raebiger et al.~\cite{Raebiger:2008} have pointed out that the net physical charge belonging to a TM atom is essentially independent of its OS and this is due to the negative-feedback charge regulation mechanism that is inherent to TM compounds~\cite{Resta:2008, Jansen:2008}. The difficulty in accurate and unambiguous determination of the OS of ions has inspired the development of novel methods. Among these, we highlight the method of Ref.~\cite{Jiang:2012} which is based on wavefunction topology and the modern theory of polarization, and the method of Ref.~\cite{Sit:2011} which is the projection-based method that uses eigenvalues of the atomic occupation matrix to determine the OS. Whereas the OS as defined in Ref.~\cite{Jiang:2012} has proven to be effective for transport processes~\cite{Pegolo:2020}, here we choose to adopt the method of Ref.~\cite{Sit:2011} which is particularly well suited for the purpose of the present work. In Sec.~S5 in SM~\cite{Note:SM:2022} we also discuss the determination of OS based on magnetic moments computed by integrating the difference between the spin-up and spin-down components of the spin-charge density over atomic spheres of varying radius centered on ions~\cite{Reed:2002}.

\begin{figure*}[t]
  \includegraphics[width=0.9\linewidth]{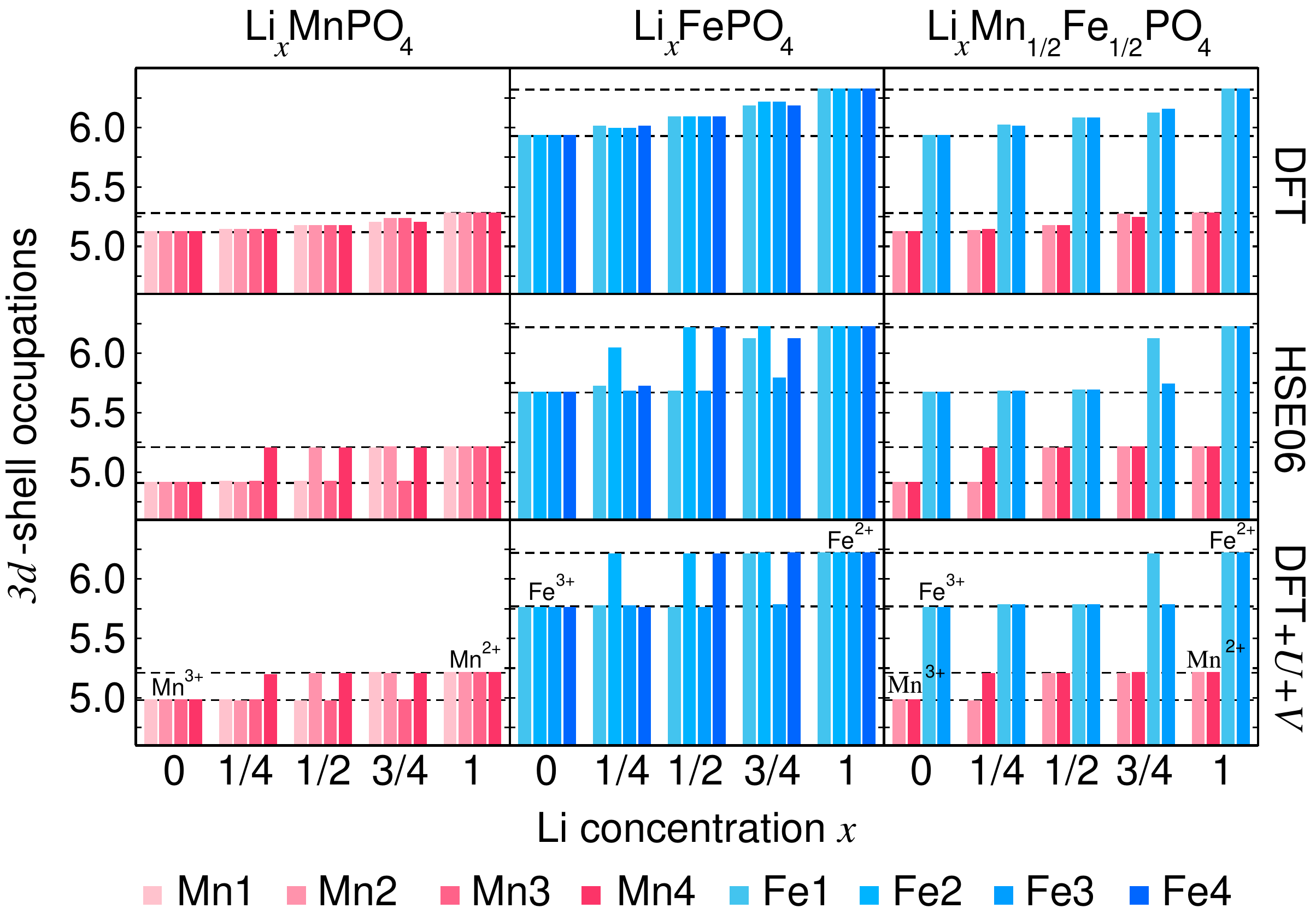}
   \caption{L\"owdin occupations of the $3d$ shell of Mn and Fe atoms in Li$_x$MnPO$_4$, Li$_x$FePO$_4$, and Li$_x$Mn$_{1/2}$Fe$_{1/2}$PO$_4$ at $x=0, 1/4, 1/2, 3/4, 1$ computed using three approaches (DFT, DFT+$U$+$V$, and HSE06). The horizontal dashed lines correspond to the L\"owdin occupations of the end elements ($x=0$ and $x=1$) with their corresponding oxidation states determined using the data in Table~\ref{tab:OS_Mn_and_Fe}. For each material there are four TM atoms, each of which is represented with a bar.}
\label{fig:occupations}
\end{figure*}

Table~\ref{tab:OS_Mn_and_Fe} reports the population analysis data for the $3d$ shell of Mn and Fe ions in Li$_x$MnPO$_4$ and Li$_x$FePO$_4$ at $x=0$ and $x=1$ computed using four approaches (DFT, DFT+$U$, DFT+$U$+$V$, and HSE06) in comparison with the occupations that can be inferred from the nominal oxidation state of the same ion. More specifically, it shows the eigenvalues of the site-diagonal ($I=J$) atomic occupation matrix $n^{I\sigma}_{mm'}$ of size $5 \times 5$ [see Eq.~\eqref{eq:occ_matrix_0}] in the spin-up ($\sigma = \uparrow$ : $\lambda_i^\uparrow$) and spin-down ($\sigma = \downarrow$ : $\lambda_i^\downarrow$) channels, L\"owdin occupations $n = \sum_{i=1}^5 (\lambda_i^\uparrow + \lambda_i^\downarrow)$, magnetic moments $m = \sum_{i=1}^5 (\lambda_i^\uparrow - \lambda_i^\downarrow)$, and the OS determined using the method of Ref.~\cite{Sit:2011}. The same analysis has been performed for Li$_x$Mn$_{1/2}$Fe$_{1/2}$PO$_4$ at $x=0$ and $1$ and is discussed in Sec.~S6 in SM~\cite{Note:SM:2022}; we do not show these results here since they are similar to those presented in Table~\ref{tab:OS_Mn_and_Fe}. As can be seen from the eigenvalues in Table~\ref{tab:OS_Mn_and_Fe}, the charge allocated on the $3d$ shell of TM ions (Fe and Mn) contains contributions from both the fully occupied $d$ orbitals (i.e. the eigenvalues that are close to 1.0~\cite{Deviation_from_one:2022} and are shown in bold) and nominally empty ones that, hybridizing with O-$2p$ states, give rise to fractional occupations. According to Ref.~~\cite{Sit:2011}, in order to determine the OS we need to count how many $d$ states are ``fully occupied''; by following this procedure and recalling the valence electronic configurations of TM atoms considered here (Mn: $3d^5 4s^2$ and Fe: $3d^6 4s^2$) we readily find that in the fully delithiated olivines ($x=0$) the OS of Mn and Fe are $+3$ while in the fully lithiated olivines ($x=1$) the OS of Mn and Fe are $+2$. This agrees well with the nominal OS shown in Table~\ref{tab:OS_Mn_and_Fe} and depicted in Fig.~\ref{fig:Nominal_OS}. In addition we find that Mn and Fe are in a high-spin state in agreement with experiments~\cite{Rousse:2003, Ofer:2012, Gnewuch:2020}. Different methods considered here give slightly different occupations of the formally empty $d$ states: For instance, in LiMnPO$_4$ the unoccupied $d$ state in the spin-up channel (corresponding to $\lambda_1^\uparrow$) features occupations in the range from 0.40 to 0.54 due to mixing with O-$2p$ ligand states, whereas much smaller filling of all $d$ states occurs in the spin-down channel. Therefore, a larger deviation from 0 of the eigenvalues indicates a stronger mixing of the unoccupied $d$ orbitals with the ligand orbitals, in accordance with the prescription of Ref.~\cite{Sit:2011}. 

Table~\ref{tab:OS_Mn_and_Fe} also contains the L\"owdin occupations $n$ and magnetic moments $m$, which are often used to determine the OS of ions. However, as we discussed earlier, these are not always appropriate descriptors of the OS: due to the hybridization of the TM orbitals with the states of their ligands, it is difficult to assign the correct number of electrons to the TM ions; moreover, the number of electrons on TM orbitals undergoes smaller changes than predicted by the nominal OS during (de-)lithiation due the negative-feedback charge regulation mechanism discussed in Ref.~\cite{Raebiger:2008}. Indeed, from Table~\ref{tab:OS_Mn_and_Fe} we can see that e.g. for FePO$_4$ and LiFePO$_4$ the nominal L\"owdin occupations are 5.0 and 6.0, respectively, while the computational predictions on average give 5.8 and 6.2 (with DFT giving the largest deviations from the nominal occupations due to SIE). Magnetic moments are also often used to determine the OS, but here we can see that these are also not appropriate quantities: the nominal magnetic moments for FePO$_4$ and LiFePO$_4$ are 5.0 and 4.0~$\mu_\mathrm{B}$, respectively, while the computational predictions on average give 4.2 and 3.7~$\mu_\mathrm{B}$ (again, the largest deviations from the nominal magnetic moments are those of DFT due to SIE). It is interesting to note that DFT+$U$+$V$ predicts the L\"owdin occupations and magnetic moments in remarkable agreement with the HSE06 ones for LiFePO$_4$, while for FePO$_4$ the DFT+$U$ results are closer to HSE06 than the DFT+$U$+$V$ ones. Similar trends are also observed for MnPO$_4$ and LiMnPO$_4$, which suggests that the TM-ligand intersite electronic interactions are slightly stronger in the fully lithiated olivines. Nevertheless, the L\"owdin occupations (and magnetic moments) are still very useful quantities for bookkeeping~\cite{Raebiger:2008}, in particular when describing the gradual (de-)lithiation process as discussed in the following.

Figure~\ref{fig:occupations} shows the L\"owdin occupations of the $3d$ shells of Mn and Fe atoms in Li$_x$MnPO$_4$, Li$_x$FePO$_4$, and Li$_x$Mn$_{1/2}$Fe$_{1/2}$PO$_4$ at $x=0, 1/4, 1/2, 3/4, 1$ computed using three approaches (DFT, DFT+$U$+$V$, and HSE06). Here we do not show the DFT+$U$ results since these are known to be less accurate than the DFT+$U$+$V$ ones in olivines e.g. for $x=1/2$~\cite{Cococcioni:2019}; in addition, the simultaneous convergence of the Hubbard $U$ parameters within DFT+$U$ and the crystal structure in a self-consistent fashion~\cite{Timrov:2021} is problematic for $x=1/4$ and $3/4$ (which requires further investigation). We stress that no convergence issues were found when using self-consistent DFT+$U$+$V$. Our main goal here is to compare the accuracy of the DFT+$U$+$V$ approach versus the well-established HSE06 one. In the case of Li$_x$MnPO$_4$, we can see that DFT+$U$+$V$ and HSE06 agree remarkably well and both show a ``digital'' change in the L\"owdin occupations: adding one Li$^+$ ion and one electron to the cathode during the lithiation process leads to changes in the occupation from 4.98 to 5.21 (and to the corresponding change in the OS from $+3$ to $+2$, see Table~\ref{tab:OS_Mn_and_Fe}) of only one Mn ion (that accepts this extra electron) while all other Mn ions remain unchanged. This process continues when we go on with the Li intercalation until eventually all Mn ions reduce from $+3$ to $+2$. Thus, these two approaches successfully describe the mixed-valence nature of the Li$_x$MnPO$_4$ compound that contains two types of Mn ion, Mn$^{3+}$ and Mn$^{2+}$, at $x=1/4, 1/2, 3/4$. In contrast, DFT fails to localize an extra electron on one of the Mn ions and as a consequence the charge density is spread out and equally distributed among all Mn ions in the system with approximately equal occupations, as can be seen in Fig.~\ref{fig:occupations}. Hence, in DFT at $x=1/4, 1/2, 3/4$ there is only one type of Mn ions whose occupations are intermediate (and progressively changing with Li content) between those of the $+2$ and $+3$ ions. 
In the case of Li$_x$FePO$_4$, our results are similar with the difference that here only DFT+$U$+$V$ shows the ``digital'' change in L\"owdin occupations while HSE06 does not manage to describe accurately the localization of electrons on Fe ions. This seems to suggest that the global mixing parameter of 0.25 of HSE06 turns out to be ineffective at describing complex electronic interactions in Li$_x$FePO$_4$, while DFT+$U$+$V$ with site-dependent self-consistent Hubbard $U$ and $V$ parameters proves capable at capturing the local chemistry (in particular, the varying amount of $3d-2p$ intersite hybridization) and the ``digital'' change in the OS of Fe ions. Finally, in Li$_x$Mn$_{1/2}$Fe$_{1/2}$PO$_4$ we find that Mn$^{3+}$ ions are the first to reduce to Mn$^{2+}$ when lithiating the compound from $x=0$ to $x=1/4$ and $x=1/2$, and only at higher Li concentrations (from $x=1/2$ to $x=1$) Fe$^{3+}$ ions reduce to Fe$^{2+}$.  Importantly, we find that both DFT+$U$+$V$ and HSE06 agree to describe the change in the L\"owdin occupations on Mn ions, while for Fe ions we again find that DFT+$U$+$V$ outpaces HSE06 in terms of accuracy; in fact, the change in the occupation of Fe-3$d$ states is not as sharp as the one obtained from the Hubbard correction. Similar trends are also observed for magnetic moments for these three materials (see Sec.~S5 in SM~\cite{Note:SM:2022}).

\subsection{Spin-resolved projected density of states}
\label{sec:results-PDOS}

In this section we analyze the spin-resolved PDOS using three approaches (DFT, DFT+$U$+$V$, and HSE06). In Fig.~\ref{fig:PDOS_LFPO} we show the spin-resolved PDOS for Li$_x$Mn$_{1/2}$Fe$_{1/2}$PO$_4$ at different concentrations of Li ($x = 0, 1/4, 1/2, 3/4, 1$) as a representative example of phospho-olivines considered in this paper, while in Sec.~S7 in the SM~\cite{Note:SM:2022} we show the spin-resolved PDOS for Li$_x$MnPO$_4$ and Li$_x$FePO$_4$.

\begin{figure*}[t]
  \includegraphics[width=0.7\linewidth]{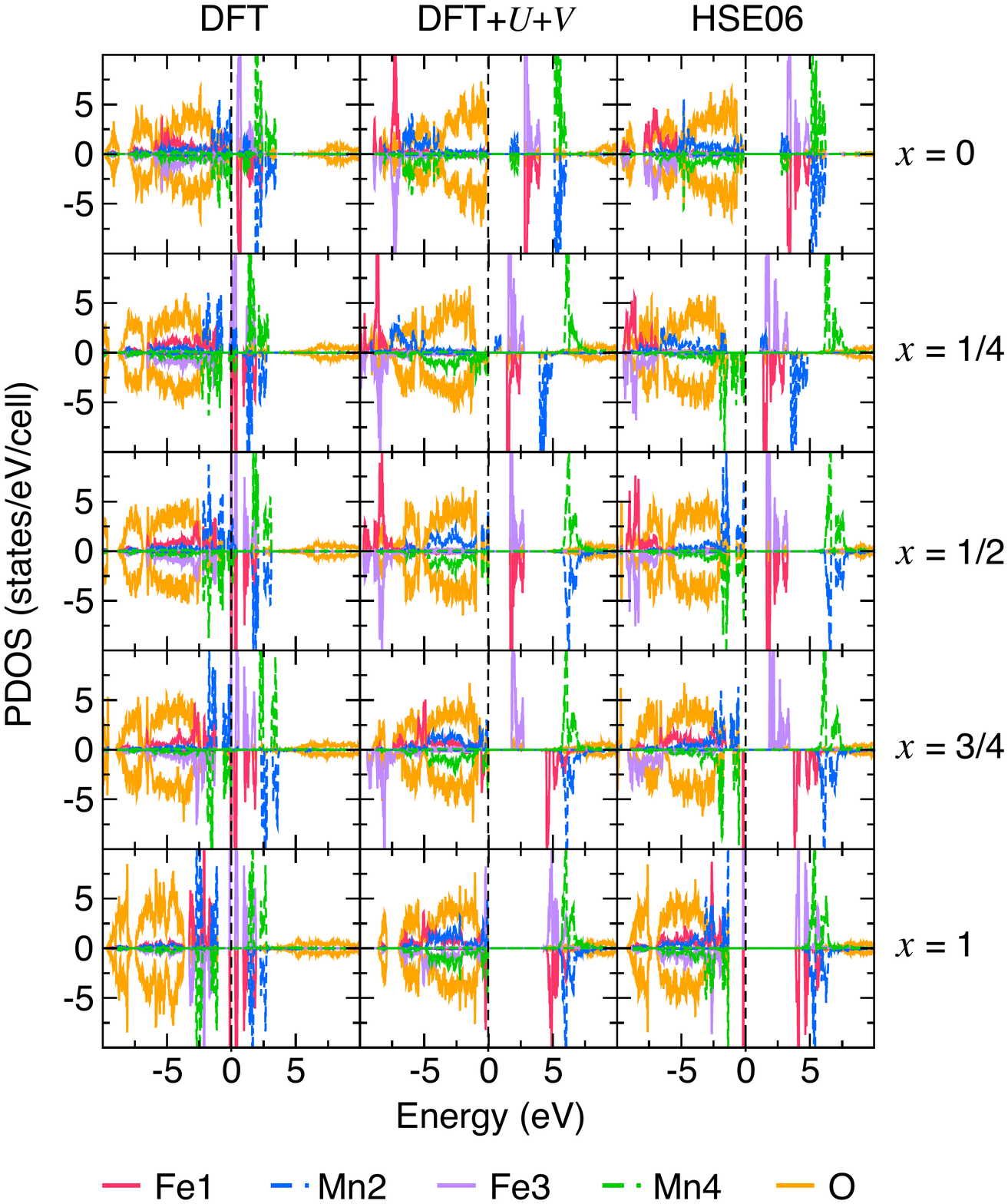}
   \caption{Spin-resolved PDOS in Li$_x$Mn$_{1/2}$Fe$_{1/2}$PO$_4$ at different concentrations of Li ($x = 0, 1/4, 1/2, 3/4, 1$) for $3d$ states of Fe1, Mn2, Fe3, Mn4 and for $2p$ states of O, computed using DFT, DFT+$U$+$V$, and HSE06. The PDOS for O-$2p$ states was obtained by summing up contributions from all O atoms in the simulation cell and it was multiplied by a factor of $1/2$ in order to have clearer comparison with the PDOS of Fe and Mn atoms. The zero of energy corresponds to the top of the valence bands in the case of insulating ground states or the Fermi level in the case of metallic ground states. The upper part of each panel corresponds to the spin-up channel, and the lower part corresponds to the spin-down channel.}
\label{fig:PDOS_LFPO}
\end{figure*}

We can see from Fig.~\ref{fig:PDOS_LFPO} that overall the PDOS computed using DFT+$U$+$V$ and HSE06 agree very well qualitatively, while the PDOS computed using DFT shows significantly different trends. More specifically, due to the overdelocalization of $d$ electrons of TM ions in DFT caused by SIE the Fe-$3d$ and Mn-$3d$ states are grouped around the Fermi level and the material exhibits spurious metallic character at $x=1/4,1/2,3/4$. Furthermore, when increasing the concentration of Li within DFT, there are no clear trends in the changes of PDOS and there is no evidence that only one TM element changes its OS from $+3$ to $+2$ (in agreement with the population analysis of Sec.~\ref{sec:results-occupations_and_OS}). Instead, the Li-donated extra electron is spread out over all Fe and Mn ions that results in approximately equal PDOS for the likewise TM elements. In contrast, both DFT+$U$+$V$ and HSE06 change drastically the PDOS compared to the DFT-based one: the material preserves its insulating character (i.e., a finite band gap) during the whole process of lithiation from $x=0$ up to $x=1$ (the reader is referred to Sec.~S8 in the SM~\cite{Note:SM:2022} for the values of band gaps). 
When changing $x$ from $0$ to $1/4$ (i.e. intercalating one Li$^+$ ion and adding one electron to the olivine cathode material), only the PDOS of one Mn ion (labeled as ``Mn4'') changes by shifting Mn-$3d$ empty states to higher energies in the spin-up channel and Mn-$3d$ occupied states closer to the top of the valence bands in the spin-down channel. Furthermore, by changing $x$ from $1/4$ to $1/2$ the PDOS of the second Mn ion (labeled as ``Mn2'') changes in the same way as the PDOS of Mn4 but mirrored with respect to the spin channels. This is in line with the fact that Mn is the first species to change its occupation when lithiating the structure from $x=0$ to $x=1/2$ as shown in Fig.~\ref{fig:occupations}. By further lithiating the cathode material from $x=1/2$ to $3/4$ and finally from $3/4$ to $1$ we can see that now the PDOS of the two Fe ions (labeled as ``Fe1'' and ``Fe3'') change by shifting Fe-$3d$ empty states to higher energies and Fe-$3d$ occupied states towards the top of the valence bands. These changes in the PDOS are consistent with the reduction of Mn$^{3+}$ to Mn$^{2+}$ and of Fe$^{3+}$ to Fe$^{2+}$ (see Sec.~\ref{sec:results-occupations_and_OS}). It is interesting to note that the occupied Fe-$3d$ states are localized and show small hybridization with O-$2p$ ligand states for $x$ in the range $0$ to $1/2$ (red and violet peaks at around $-8$~eV) while, for $x$ from $1/2$ to $1$, they move up in energy thus overlapping with the O-$2p$ states (from $-7$ to 0~eV) and becoming more dispersive. 
This latter effect is well captured both in HSE06 and DFT+$U$+$V$ since both describe the intersite electronic interactions and not only the localization of $d$ electrons.

It is instructive to highlight the differences in the PDOS computed using DFT+$U$+$V$ and HSE06. While both methods show changes in the character of the top of the valence bands when going from $x=0$ to $1$, the fine details are different. At $x=0$, DFT+$U$+$V$ shows that the top of the valence bands is strongly dominated by the O-$2p$ states while HSE06 predicts that the top of the valence bands is more of a mixed nature due to the hybridization between Mn-$3d$ and O-$2p$ states. In addition, in DFT+$U$+$V$ we can see a clearer energy separation between the Fe-$3d$ and Mn-$3d$ empty states at $x=0$, while within HSE06 these states are closer in energy. At $x=1/4$ and $x=1/2$, both methods show that the top of the valence bands is dominated by the Mn-$3d$ states, although in HSE06 the intensity of these states is much stronger than in DFT+$U$+$V$. Finally, at $x=3/4$ and $1$ these two methods give different predictions for the character of the top of the valence bands. In HSE06 at $x=1$, Fe-$3d$ states are the highest occupied states while Mn-$3d$ states lie deeper in energy and there is an energy gap between these two sets of states. In contrast, in DFT+$U$+$V$ at $x=1$ there is no gap between the Fe-$3d$ and Mn-$3d$ occupied states, and all these states overlap in energy and thus the top of the valence bands is predominantly of the Fe-$3d$ and Mn-$3d$ character. To the best of our knowledge there is no experimental data from photoemission and x-ray absorption spectroscopy measurements, so it is not possible to establish which method gives a more accurate description of the electronic structure of Li$_x$Mn$_{1/2}$Fe$_{1/2}$PO$_4$. However, the fact that DFT+$U$+$V$ can capture the ``digital'' change of L\"owdin occupations (especially for Fe ions) upon the lithiation of olivines (as shown in Sec.~\ref{sec:results-occupations_and_OS}) suggests that the PDOS from DFT+$U$+$V$ is probably more reliable than that from HSE06. Further investigations are required in order to shed more light on this issue. But the overall agreement between trends in the PDOS computed within DFT+$U$+$V$ and HSE06 proves that these two methods - despite having very different mathematical formulations and theoretical background (see Sec.~\ref{sec:methods}) - yield on average similar predictions of the electronic structure of phospho-olivines.

\subsection{Lithium intercalation voltages}
\label{sec:results-Voltages}

\begin{figure*}[t]
  \includegraphics[width=0.8\linewidth]{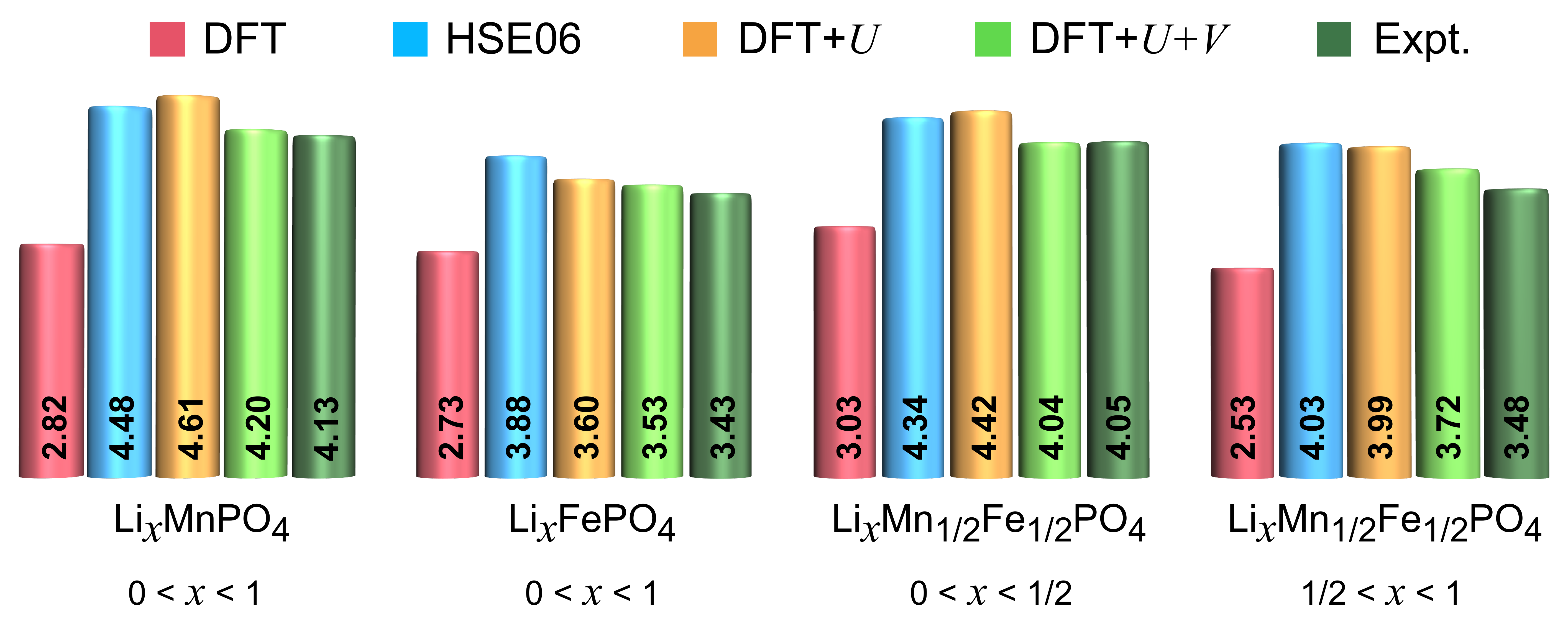}
   \caption{Voltages vs. Li/Li$^+$ (in V) for Li$_x$MnPO$_4$, Li$_x$FePO$_4$, and Li$_x$Mn$_{1/2}$Fe$_{1/2}$PO$_4$ computed using DFT, HSE06, DFT+$U$, and DFT+$U$+$V$ with $U$ and $V$ determined from first-principles. The experimental data is from Refs.~\cite{Kobayashi:2009, Muraliganth:2010}.}
\label{fig:voltages}
\end{figure*}

The topotactic Li intercalation voltages can be computed using the fundamental thermodynamic definition~\cite{Aydinol:1997, Cococcioni:2019}:
\begin{equation}
    \Phi = - \frac{E(\mathrm{Li}_{x_2}\mathrm{S}) - E(\mathrm{Li}_{x_1}\mathrm{S}) - (x_2 - x_1)E(\mathrm{Li})}{(x_2 - x_1) e} ,
    \label{eq:voltage}
\end{equation}
where S is introduced for the sake of shorthand notation and it denotes e.g. MnPO$_4$ for Li$_x$MnPO$_4$ and similarly for other cathode materials considered in this paper. Here, $\Phi$ is the voltage, $e$ is the electronic charge, $x_1$ and $x_2$ are the concentrations of Li and they take values between 0 and 1 in this study, and $E$ is the total energy per formula unit. It is important to remark that $E(\mathrm{Li})$ is the total energy of bulk Li computed at the level of standard DFT (PBEsol functional) while $E(\mathrm{Li}_{x_1}\mathrm{S})$ and $E(\mathrm{Li}_{x_2}\mathrm{S})$ are computed using the four approaches considered in this work: DFT, HSE06, DFT+$U$, and DFT+$U$+$V$ ($U$ and $V$ are computed self-consistently individually for each structure~\cite{Cococcioni:2019}). We note that entropic and pressure-volume effects are neglected when computing $\Phi$ since these are known to not significantly impact average Li intercalation voltages~\cite{Aydinol:1997b}.

Figure~\ref{fig:voltages} shows a comparison between computed voltages and experimental ones from Refs.~\cite{Kobayashi:2009, Muraliganth:2010}. For Li$_x$MnPO$_4$ and Li$_x$FePO$_4$ we compute the average voltages in the range $0<x<1$ (thus $x_1=0$ and $x_2=1$) since experimentally it is known that there is only one plateau in the voltage profile~\cite{Muraliganth:2010}. In contrast, for Li$_x$Mn$_{1/2}$Fe$_{1/2}$PO$_4$ we compute the average voltages in two ranges of $x$, $0<x<1/2$ and $1/2<x<1$, which correspond to the two plateaus observed experimentally in the voltage profile~\cite{Muraliganth:2010}. As discussed in previous sections, $0<x<1/2$ corresponds to the reduction of Mn ions and hence the voltage is similar to that of Li$_x$MnPO$_4$, while $1/2<x<1$ corresponds to the reduction of Fe ions with a voltage that is similar to that of Li$_x$FePO$_4$. Experimentally it is known that the mixing of TM cations creates shifts in redox potentials: the voltage of the Mn$^{2+/3+}$ couple is decreased by $\sim0.08$~V while the voltage of the Fe$^{2+/3+}$ couple is increased by $\sim0.05$~V when going from the pristine end members (Li$_x$MnPO$_4$ and Li$_x$FePO$_4$) to the mixed TM olivine (Li$_x$Mn$_{1/2}$Fe$_{1/2}$PO$_4$)~\cite{Kobayashi:2009, Muraliganth:2010}. The shifts in redox potentials were also observed in previous DFT+$U$-based calculations~\cite{Malik:2009, Snydacker:2016, Loftager:2019} and were attributed to changes in the TM$-$O bond lengths~\cite{Kobayashi:2009, Muraliganth:2010, Loftager:2019} or strain~\cite{Snydacker:2016}.

As can be seen from Fig.~\ref{fig:voltages}, standard DFT largely underestimates the voltages (on average $22-31$\% off with respect to the experiments). This demonstrates that the energetics is strongly affected by the delocalization of TM $d$ electrons due to the strong SIE inherent to xc functionals (such as e.g. PBEsol). HSE06 alleviates these errors partially and improves the energetics, however the resulting voltages are overestimated by $6-15$\%. It is worth noting that our HSE06-based voltages are significantly higher than those of Ref.~\cite{Chevrier:2010} that reports 3.87~V for Li$_x$MnPO$_4$ and 3.33~V for Li$_x$FePO$_4$ using HSE06. These discrepancies are likely due to various differences in computational details (different pseudopotentials, screening parameter $\omega$, kinetic-energy cutoff, $\mathbf{k}$ points sampling, etc.) and different geometries (here we use the DFT+$U$+$V$ geometry for HSE06 calculations while in Ref.~\cite{Chevrier:2010} the HSE06-optimized geometry was used). We recall that in HSE06 the amount of Fock exchange is fixed to 25\%, and it is quite a common practice to adjust this percentage by reproducing e.g. the experimental band gaps, which as a byproduct can lead to more accurate intercalation voltages~\cite{Seo:2015}. However, the semiempirical adjustment of the amount of Fock exchange often relies on high-resolution experimental data, which are not always available.

Figure~\ref{fig:voltages} shows that DFT+$U$ manifests different trends with respect to HSE06 voltages depending on the material and the range of $x$ considered: compared to HSE06 voltages it achieves somewhat higher values for Li$_x$MnPO$_4$, but lower ones for Li$_x$FePO$_4$ (for which voltages result closer to the experimental value). Overall, DFT+$U$ voltages are scattered over a wider range ($3-14$\%) around the experimental values than those obtained from HSE06. It is useful to remark that our DFT+$U$ voltages for Li$_x$MnPO$_4$ and Li$_x$FePO$_4$ are in better agreement with the experimental ones than those of Ref.~\cite{Cococcioni:2019}; as was pointed out in Ref.~\cite{Timrov:2021}, this is a consequence of the difference in the values of $U$, and of the consistent calculation of forces and stresses using the orthogonalized atomic Hubbard projectors [see Eq.~\eqref{eq:OAO_def}] that has significantly refined the prediction of the equilibrium crystal structure in this work. Finally, DFT+$U$+$V$ gives the most accurate predictions of voltages compared to all other methods considered in this work. More specifically, the average deviation of DFT+$U$+$V$ voltages from the experimental values is in the $1-7$\% range; leaving aside the Li$_x$Mn$_{1/2}$Fe$_{1/2}$PO$_4$ case with $1/2<x<1$, the average deviation is $1-2$\% which is remarkable given the fact that the DFT+$U$+$V$ calculations are fully first-principles with no fitting or adjusted parameters. This finding demonstrates that the accuracy of the DFT+$U$+$V$ approach with $U$ and $V$ computed using linear-response theory~\cite{Cococcioni:2005, Timrov:2018} in a self-consistent fashion~\cite{Cococcioni:2019, Timrov:2021} is satisfactory for predictive simulations of olivine-type cathode materials. Regarding the redox potential shifts of the two plateaus of Li$_x$Mn$_{1/2}$Fe$_{1/2}$PO$_4$ compared to Li$_x$MnPO$_4$ and Li$_x$FePO$_4$, within DFT+$U$+$V$ we find values of 0.16 and 0.19~V for the Mn$^{2+/3+}$ and Fe$^{2+/3+}$ couples, respectively. These redox potential shifts are similar to those obtained within HSE06, namely 0.14 and 0.15~V for the Mn$^{2+/3+}$ and Fe$^{2+/3+}$ couples, respectively. Therefore, both DFT+$U$+$V$ and HSE06 overestimate the experimental redox potential shifts. At the same time we observe changes in the Mn$-$O and Fe$-$O bond lengths in the mixed TM olivine compared to the pristine end members (see Table~S2 of SM~\cite{Note:SM:2022}), in consistency with the hypothesis of Refs.~\cite{Muraliganth:2010, Kobayashi:2009, Loftager:2019} that these might be responsible for the redox potential shifts.

These promising results and observations motivate investigations of other classes of cathode materials using the extended Hubbard functional, and the work in this direction is in progress. Furthermore, the predictive power of DFT+$U$+$V$ might help to obtain further insights on still problematic aspects of the considered systems, e.g. the asymmetric charge-discharge behavior of Li$_x$FePO$_4$ possibly promoted by the existence of a hidden two-step phase transition via a metastable phase~\cite{Koyama:2017}.

\subsection{General remarks}
\label{sec:discussion}

In the previous sections we have shown that DFT+$U$+$V$ is a powerful tool for the accurate description of the structural, electronic, magnetic, and electrochemical properties of phospho-olivines. It is useful to provide general remarks about this approach compared to state-of-the-art approaches that are currently used.

\textit{Computational cost.} DFT+$U$+$V$ is only marginally more expensive than DFT+$U$ when the parameters $U$ and $V$ are known, and both these methods are only slightly more demanding than plain DFT. However, the cost of computing $U$ and $V$ using DFPT is an order-of-magnitude larger (with some prefactor that depends on the number of symmetries, number of nonequivalent atoms of the same type, etc.) than ground-state DFT calculations. Therefore, the cost of the DFT+$U$+$V$ calculation itself is negligible compared to the cost of the first-principles determination of the Hubbard parameters. However, the values of $U$ and $V$ can be easily machine learned based on the DFPT data, as we will argue in a future publication. Compared to the computational effort associated with hybrid functionals (e.g., HSE06), that required by the self-consistent DFPT evaluation of the Hubbard parameters is still lower, especially when the number of atoms is large (greater than $\sim 10$~\cite{KirchnerHall:2021}). We recall that the structural optimizations using HSE06 is extremely expensive for systems containing several tens of atoms (like for phospho-olivines) and hence it was not performed in this work (although it was reported in other works, e.g. in Ref.~\cite{Chevrier:2010}). Conversely, structural optimization using DFT+$U$+$V$ are absolutely affordable and the main cost comes from the evaluation of Hubbard forces and stresses using L\"owdin-orthogonalized atomic orbitals~\cite{Timrov:2020b}. We note that the structural optimization using DFT+$U$+$V$ is inherently incorporated in the self-consistent protocol of the evaluation of the Hubbard parameters that brings the system to the global minimum~\cite{Timrov:2021}. Therefore, overall the self-consistent DFT+$U$+$V$ approach is much more affordable than HSE06 although obviously more expensive than DFT+$U$ with empirical $U$ parameters.

\textit{Dependence on the availability of the experimental data.} The strength of the DFT+$U$+$V$ approach used here is that it is ``parameter-free'' in the sense that $U$ and $V$ are computed from first principles without relying on any experimental data. This makes this approach predictive for novel materials for which experimental data are not available and allows us to capture the dependence of the Hubbard parameters, e.g., on the local chemical environment and on the OS. As for what concerns hybrids, when tuning the fraction of Fock exchange is necessary to improve their predictivity an empirical strategy can also be adopted, presenting the same disadvantages as for empirically-tuned Hubbard-corrected functionals. First-principles calculations of these parameters have also become increasingly popular in recent years~\cite{Skone:2014, Skone:2016, Bischoff:2019, Kronik:2012, Wing:2021, Lorke:2020}: however, they tend to further increase the already significant computational costs. 

\textit{Generalizations and limitations.} The DFT+$U$+$V$ framework is very general, it can be used with any xc functional, e.g. PBE~\cite{Perdew:1996}, PBEsol~\cite{Perdew:2008}, SCAN~\cite{Sun:2015}, rSCAN~\cite{Bartok:2019}, r$^2$SCAN~\cite{Furness:2020}, etc. Since SCAN and its flavors are gaining more and more popularity in the community, it would be very useful and important to generalize DFT+$U$+$V$ to meta-GGAs. In practice, though, this requires, first of all, the availability of the meta-GGA pseudopotentials, and, secondly, the generalization of DFPT to meta-GGAs for a consistent evaluation of the Hubbard interactions. As for what concerns limitations, currently our formulation of DFT+$U$+$V$ does not include the Hund's $J$ corrections that are known to be important in some classes of materials~\cite{Himmetoglu:2011}. Moreover, DFT+$U$+$V$ is a mean-field approach based on a single Slater determinant, hence systems for which the multireference nature of the wave function is important are beyond reach for the current formulation of DFT+$U$+$V$. Finally, as mentioned earlier, the simultaneous convergence of the Hubbard $U$ parameters within DFT+$U$ and the crystal structure in a self-consistent fashion~\cite{Timrov:2021} is problematic for some fractional concentrations of Li that might be due to the missing derivatives of the Hubbard parameters with respect to atomic positions when computing Hubbard forces~\cite{Kulik:2011b}.

\textit{Data set and databases.} In the present work we consider only three examples from the same family of cathode materials (containing only two different TM ions, Mn and Fe). Therefore, our work is by no means conclusive and the accuracy of DFT+$U$+$V$ for other classes of cathode materials (e.g. layered, spinel, etc.) has to be verified thoroughly and with care. However, the promising results presented in this work for phospho-olivines are very encouraging and the whole computational DFT+$U$+$V$ framework is indeed very robust. Work is in progress for the development of automated DFT+$U$+$V$ workflows for the high-throughput calculations for cathode materials using AiiDA~\cite{Pizzi:2016, Huber:2020}, which would allow us to generate large databases of cathode materials' properties and benchmark them versus data obtained using state-of-the-art methods and data from experiments.

\section{CONCLUSIONS}
\label{sec:conclusions}

We have presented the first comparative study (using DFT, DFT+$U$, DFT+$U$+$V$, and HSE06) of the electronic properties and the energetics of lithium intercalation in
representative 
phospho-olivine cathode materials: Li$_x$MnPO$_4$, Li$_x$FePO$_4$, and Li$_x$Mn$_{1/2}$Fe$_{1/2}$PO$_4$ ($x=0, 1/4, 1/2, 3/4, 1$). In DFT+$U$ and DFT+$U$+$V$, the Hubbard parameters $U$ and $V$ have been computed from first principles using density-functional perturbation theory, without any need for adjustments or {\em ad hoc} fitting of the model. 

By determining the oxidation state of TM ions using the projection-based method of Ref.~\cite{Sit:2011}, we were able to analyse the change in L\"owdin occupations of the $d$ manifolds during the lithiation process. We have found that DFT fails to 
account for the onset of disproportionation of the TM atoms along the intermediates of the lithiation process. In contrast, DFT+$U$+$V$ correctly predicts the ``digital'' change of L\"owdin occupations upon Li intercalation (only one TM ion changes its oxidation state from $+3$ to $+2$ for each Li ion added) in all materials studied here. For comparison, HSE06 shows the ``digital'' change in occupations for Li$_x$MnPO$_4$ but it fails to do so for Li$_x$FePO$_4$ at $x=1/4$ and $3/4$ and for Li$_x$Mn$_{1/2}$Fe$_{1/2}$PO$_4$ at $x=3/4$. Furthermore, the investigation of the electronic structure has revealed that both DFT+$U$+$V$ and HSE06 qualitatively show similar trends in the spin-resolved projected density of states, while DFT fails dramatically due to strong self-interactions errors. 

Finally, the computed intercalation voltages are greatly underestimated within DFT, whereas HSE06 brings voltages closer to the experimental values, albeit with a slight systematic overestimation. On the other hand, while DFT+$U$ is on average only slightly worse than HSE06, DFT+$U$+$V$ outperforms HSE06 in terms of accuracy, achieving voltages in very good agreement with experiments. These findings motivate the investigation of the electrochemical properties of other classes of cathode materials (e.g. layered, spinel, etc.) using DFT+$U$+$V$, and work is underway along these paths. Finally, this paves opens the way for a reliable and fully first-principles design and characterization of novel cathode materials with affordable computational costs and high level of accuracy.

\section*{ACKNOWLEDGEMENTS}

We acknowledge support from the Swiss National Science Foundation (SNSF), through Grant No.~200021-179138, and its National Centre of Competence in Research (NCCR) MARVEL. 
F.A. acknowledges the European H2020 Intersect project, Grant No.~814487.
Computer time was provided by the Swiss National Supercomputing Centre (CSCS) under project No.~s1073.


%

\clearpage
\clearpage 
\setcounter{page}{1}
\renewcommand{\thetable}{S\arabic{table}}  
\setcounter{table}{0}
\renewcommand{\thefigure}{S\arabic{figure}}
\setcounter{figure}{0}
\renewcommand{\thesection}{S\arabic{section}}
\setcounter{section}{0}
\renewcommand{\theequation}{S\arabic{equation}}
\setcounter{equation}{0}
\onecolumngrid

\begin{center}
\textbf{\large Supplemental Material for}
\vskip 0.2 cm
\textbf{\large ``Accurate electronic properties and intercalation voltages of olivine-type Li-ion cathode materials from extended Hubbard functionals''}
\vskip 0.2cm
Iurii Timrov, Francesco Aquilante, Matteo Cococcioni, and Nicola Marzari
\end{center}

\maketitle

\section{Technical details}
\label{sec:technical_details}

All calculations were performed using the plane-wave (PW) pseudopotential implementation of DFT contained in the \textsc{Quantum ESPRESSO} distribution~\cite{Giannozzi:2009SM, Giannozzi:2017SM, Giannozzi:2020SM}. 

We have used the exchange-correlation functional constructed using $\sigma$-GGA with the PBEsol prescription~\cite{Perdew:2008SM}. For DFT, DFT+$U$, and DFT+$U$+$V$ calculations we have used pseudopotentials (PPs) from the SSSP library~v1.1 (precision)~\cite{prandini2018precisionSM, MaterialsCloudSM}, which are either ultrasoft (US) or projector-augmented-wave (PAW): For manganese we have used \texttt{mn\_pbesol\_v1.5.uspp.F.UPF} from the GBRV v1.5 library~\cite{Garrity:2014SM}, for iron and oxygen \texttt{Fe.pbesol-spn-kjpaw\_psl.0.2.1.UPF} and \texttt{O.pbesol-n-kjpaw\_psl.0.1.UPF} from the Pslibrary v0.3.1~\cite{Kucukbenli:2014SM}, for phosphorus \texttt{P.pbesol-n-rrkjus\_psl.1.0.0.UPF} from the Pslibrary v1.0.0~\cite{DalCorso:2014SM}, and for lithium \texttt{li\_pbesol\_v1.4.uspp.F.UPF} from the GBRV v1.4 library~\cite{Garrity:2014SM}. For HSE06 calculations we have used norm-conserving PPs from the \textsc{PseudoDojo} library~\cite{vanSetten:2018SM}.

To construct the Hubbard projector functions $\varphi^I_m(\mathbf{r})$ [see Eq.~(9) in the main text] we have used atomic orbitals which are orthogonalized using L\"owdin's method~\cite{Lowdin:1950SM, Mayer:2002SM}. Structural optimizations using DFT+$U$ and DFT+$U$+$V$ are performed using orthogonalized atomic orbitals as described in detail in Ref.~\cite{Timrov:2020bSM}. The Brillouin zone was sampled using the uniform $\Gamma$-centered $\mathbf{k}$ point mesh of size $5 \times 8 \times 9$. Kohn-Sham (KS) wavefunctions and potentials are expanded in PWs up to a kinetic-energy cutoff of 90 and 1080~Ry, respectively, for structural optimization. The crystal structure was optimized using the Broyden-Fletcher-Goldfarb-Shanno (BFGS) algorithm~\cite{Fletcher:1987SM}, with a convergence threshold for the total energy of $10^{-6}$~Ry, for forces of $10^{-5}$~Ry/Bohr, and for pressure of 0.5~Kbar. For the metallic ground states (that appear at the DFT level of theory and intermediate Li concentrations), we have used the Marzari-Vanderbilt smearing method~\cite{Marzari:1999SM} with a broadening parameter of 0.01~Ry. For the formation energy calculations of Li$_x$FePO$_4$ and Li$_x$Mn$_{1/2}$Fe$_{1/2}$PO$_4$ we used tighter parametrization in order to ensure the convergence of the results: we used a $8 \times 10 \times 12$ uniform $\Gamma$-centered $\mathbf{k}$ point mesh and the kinetic-energy cutoff of 120 and 1440~Ry for the KS wavefunctions and potentials, respectively.

The DFPT calculations of Hubbard parameters are performed using the \textsc{HP} code~\cite{Timrov:2022SM} of \textsc{Quantum ESPRESSO} using the uniform $\Gamma$-centered $\mathbf{k}$ and $\mathbf{q}$ point meshes of size $3 \times 4 \times 5$ and $1 \times 2 \times 3$, respectively, which give an accuracy of 0.01~eV for the computed values of $U$ and $V$. The KS wavefunctions and potentials are expanded in PWs up to a kinetic-energy cutoff of 65 and 780~Ry, respectively, for calculation of Hubbard parameters. The linear-response KS equations of DFPT are solved using the conjugate-gradient algorithm~\cite{Payne:1992SM} and the mixing scheme of Ref.~\cite{Johnson:1988SM} for the response potential to speed up convergence.

The HSE06 (PBE + 25\% short-range Fock) calculations are performed using the uniform $\Gamma$-centered $\mathbf{k}$ and $\mathbf{q}$ point meshes of size $6 \times 8 \times 10$ and $3 \times 4 \times 5$, respectively. We recall here that the use of the $\mathbf{q}$ point mesh is different in HSE06 and in DFPT: in the former case, the $\mathbf{q}$ mesh is a coarsened $\mathbf{k}$ mesh that is used to reduce the computational cost of the HSE06 calculations; in the latter case the $\mathbf{q}$ points represent the wavevectors of monochromatic perturbations that are applied to the system to study its (linear) response~\cite{Timrov:2018SM}. The $\mathbf{q} \rightarrow 0$ limit in HSE06 was treated using the Gygi-Baldereschi scheme~\cite{Gygi:1986SM}. The KS wavefunctions and potentials are expanded in PWs up to a kinetic-energy cutoff of 80 and 320~Ry, respectively, while the exact-exchange (Fock) term was expanded in PWs up to 80~Ry. All HSE06 calculations were performed using the DFT+$U$+$V$ optimized geometry (since the HSE06 structural optimizations are computationally too expensive for phospho-olivines).

Bulk Li is modelled at the DFT-PBEsol level using the \textit{bcc} unit cell with one Li atom at the origin. The optimized lattice parameter is 3.436~\AA. The Brillouin zone was sampled using the uniform $\Gamma$-centered $\mathbf{k}$ point mesh of size $10 \times 10 \times 10$, and we have used the Marzari-Vanderbilt smearing method~\cite{Marzari:1999SM} with a broadening parameter of 0.02~Ry. The KS wavefunctions and potentials are expanded in PWs up to a kinetic-energy cutoff of 65 and 780~Ry, respectively.

The spin-resolved projected density of states (PDOS) within all approaches considered here was computed using a $\Gamma$-centered $\mathbf{k}$ point mesh of size $6 \times 8 \times 10$, with the Gaussian smearing and a broadening parameter of $10^{-3}$~Ry.

The data used to produce the results of this work are available in the Materials Cloud Archive~\cite{MaterialsCloudArchive2022SM}.

\section{Hubbard parameters}
\label{sec:hub_param}

Table~\ref{tab:hub_param} contains the values of onsite $U$ and intersite $V$ Hubbard parameters for Mn($3d$) and Fe($3d$) states in phospho-olivines considered here, which were computed in the framework of DFT+$U$+$V$ self-consistently using DFPT~\cite{Timrov:2018SM, Timrov:2021SM} with orthogonalized atomic orbitals as Hubbard projectors, as described in Sec.~II.C of the main text. We have also computed the onsite Hubbard $U$ in the framework of DFT+$U$ -- these values are smaller than $U$'s reported in Table~\ref{tab:hub_param} due to differences in the electronic screening~\cite{Cococcioni:2019SM}. More specifically, within DFT+$U$ for Li$_x$MnPO$_4$ we obtain $U$(Mn)=6.19~eV for $x=0$ and $U$(Mn)=4.30~eV for $x=1$; for Li$_x$FePO$_4$, Hubbard $U$(Fe)=5.00~eV for $x=0$ and $U$(Fe)=4.99~eV for $x=1$; and, for Li$_x$Mn$_{1/2}$Fe$_{1/2}$PO$_4$, $U$(Fe)=5.01~eV and $U$(Mn)=6.23~eV for $x=0$, and $U$(Fe)=4.97~eV and $U$(Mn)=4.32~eV for $x=1$. 

\begin{table*}[h!]
\centering
\begin{tabular}{c|c|c|c|c|c|c}
\hline\hline
Material & \parbox{1cm}{$x$}  & \parbox{1cm}{HP} &  \parbox{2cm}{Mn1} & \parbox{2cm}{Mn2} & \parbox{2cm}{Mn3} & \parbox{2cm}{Mn4} \\ 
\hline
\parbox[t]{7mm}{\multirow{10}{*}{\rotatebox[origin=c]{90}{Li$_x$MnPO$_4$}}} 
& \multirow{2}{*}{0}    & $U$ &  6.26     & 6.26      & 6.26      & 6.26      \\
&                       & $V$ & 0.54-1.07 & 0.54-1.07 & 0.54-1.07 & 0.54-1.07 \\
\cline{2-7}
& \multirow{2}{*}{1/4}  & $U$ &  6.26     & 6.25      & 6.67      & 5.44      \\
&                       & $V$ & 0.40-1.01 & 0.46-1.05 & 0.54-1.11 & 0.39-1.08 \\
\cline{2-7}
& \multirow{2}{*}{1/2}  & $U$ &  6.42     & 4.95      & 6.41      & 4.94      \\
&                       & $V$ & 0.34-1.01 & 0.38-0.96 & 0.34-1.01 & 0.38-0.96 \\ 
\cline{2-7}
& \multirow{2}{*}{3/4}  & $U$ &  4.67     & 4.64      & 6.58      & 4.98      \\
&                       & $V$ & 0.48-0.72 & 0.31-0.91 & 0.33-1.02 & 0.41-0.79 \\
\cline{2-7}
& \multirow{2}{*}{1}    & $U$ &  4.56     & 4.56      & 4.56      & 4.56      \\
&                       & $V$ & 0.42-0.78 & 0.42-0.78 & 0.42-0.78 & 0.42-0.78 \\
\hline\hline
Material & \parbox{1cm}{$x$}  & \parbox{1cm}{HP} &  \parbox{2cm}{Fe1} & \parbox{2cm}{Fe2} & \parbox{2cm}{Fe3} & \parbox{2cm}{Fe4} \\ 
\hline
\parbox[t]{7mm}{\multirow{10}{*}{\rotatebox[origin=c]{90}{Li$_x$FePO$_4$}}}
& \multirow{2}{*}{0}    & $U$ &  5.43     & 5.43      & 5.43      & 5.43      \\
&                       & $V$ & 0.60-1.08 & 0.60-1.08 & 0.60-1.08 & 0.60-1.08 \\
\cline{2-7}
& \multirow{2}{*}{1/4}  & $U$ &  5.39     & 5.74      & 5.44       & 5.40      \\
&                       & $V$ & 0.53-1.01 & 0.50-1.21 & 0.61-1.09 & 0.44-0.96 \\
\cline{2-7}
& \multirow{2}{*}{1/2}  & $U$ &  5.37     & 5.58      & 5.37      & 5.58      \\
&                       & $V$ & 0.48-0.92 & 0.43-0.93 & 0.48-0.92 & 0.43-0.93 \\ 
\cline{2-7}
& \multirow{2}{*}{3/4}  & $U$ &  5.65     & 5.41      & 5.38      & 5.31      \\
&                       & $V$ & 0.49-1.01 & 0.30-0.99 & 0.48-0.93 & 0.39-0.88 \\
\cline{2-7}
& \multirow{2}{*}{1}    & $U$ &  5.29     & 5.29      & 5.29      & 5.29      \\
&                       & $V$ & 0.42-0.90 & 0.42-0.90 & 0.42-0.90 & 0.42-0.90 \\
\hline\hline
Material & \parbox{1cm}{$x$}  & \parbox{1cm}{HP} &  \parbox{2cm}{Fe1} & \parbox{2cm}{Fe3} & \parbox{2cm}{Mn2} & \parbox{2cm}{Mn4} \\ 
\hline
\parbox[t]{7mm}{\multirow{10}{*}{\rotatebox[origin=c]{90}{Li$_x$Mn$_{1/2}$Fe$_{1/2}$PO$_4$}}}
& \multirow{2}{*}{0}    & $U$ &  5.43     & 5.43      & 6.27      & 6.27      \\
&                       & $V$ & 0.60-1.12 & 0.60-1.12 & 0.55-1.05 & 0.55-1.05 \\
\cline{2-7}
& \multirow{2}{*}{1/4}  & $U$ &  5.50     & 5.40      & 6.24       & 5.01      \\
&                       & $V$ & 0.51-1.12 & 0.60-1.11 & 0.43-1.00 & 0.35-0.96 \\
\cline{2-7}
& \multirow{2}{*}{1/2}  & $U$ &  5.44     & 5.44      & 4.81      & 4.81      \\
&                       & $V$ & 0.54-1.06 & 0.54-1.06 & 0.28-0.91 & 0.28-0.91 \\ 
\cline{2-7}
& \multirow{2}{*}{3/4}  & $U$ &  5.59     & 5.42      & 4.79      & 4.59      \\
&                       & $V$ & 0.48-0.93 & 0.48-0.94 & 0.33-0.91 & 0.41-0.76 \\
\cline{2-7}
& \multirow{2}{*}{1}    & $U$ &  5.28     & 5.28      & 4.58      & 4.58      \\
&                       & $V$ & 0.41-0.89 & 0.41-0.89 & 0.42-0.80 & 0.42-0.80 \\
\hline\hline
\end{tabular}
\caption{Self-consistent Hubbard parameters (HP) in eV computed using DFPT in the DFT+$U$+$V$ framework for Mn($3d$) states in Li$_x$MnPO$_4$, for Fe($3d$) states in Li$_x$FePO$_4$, and for Mn($3d$) and Fe($3d$) in Li$_x$Mn$_{1/2}$Fe$_{1/2}$PO$_4$ for $x=0,1/4,1/2,3/4,1$.}
\label{tab:hub_param}
\end{table*}

From Table~\ref{tab:hub_param} we can see how the values of Hubbard parameters change upon lithiation of the phospho-olivines. The general trend when going from $x=0$ to $x=1$ is that the values of Hubbard parameters decrease. Namely, Hubbard $U$ decreases because the $3d$ manifolds of TM ions acquire an extra electron due to the insertion of Li. This suggests that a larger number of electrons on the same ion promotes the screening of the effective Hubbard interactions thus leading to lower values. Also intersite Hubbard $V$ values decrease when going from $x=0$ to $x=1$. This latter fact can probably be related to the structural expansion (unit cell volume and TM--O bond length, see Fig.~\ref{fig:lat_param}) that the system undergoes upon lithiation; the longer the bond between two atoms, the smaller the intersite Hubbard interaction $V$.

It is useful to make a comparison of the Hubbard $U$ and $V$ values reported in Table~\ref{tab:hub_param} with those in Ref.~\cite{Cococcioni:2019SM}. In general, we find that there is a good agreement between these two studies, with largest differences for the Hubbard parameters being $\sim 0.3$~eV in the DFT+$U$+$V$ framework. However, in the DFT+$U$ framework the largest difference in the Hubbard $U$ is $\sim 2$~eV. These differences stem from the fact that in this paper we consistently use orthogonalized atomic orbitals for the calculation of Hubbard parameters \textit{and} for the structural optimization (i.e. to compute Hubbard forces and stresses~\cite{Timrov:2020bSM}), whereas in Ref.~\cite{Cococcioni:2019SM} the orthogonalized atomic orbitals were used for the calculation of Hubbard parameters while structural optimizations were based on nonorthogonalized atomic orbitals (since at that time there was no implementation of Hubbard energy derivatives with orthogonal atomic basis set). This highlights the importance of consistency in using the same type of Hubbard projector functions across all calculations. 

Finally, it is informative to comment about a change in the Hubbard parameters when gradually changing the concentration of Li. We stress that Hubbard parameters are site-dependent quantities, i.e. they are \textit{not} global averaged values that are the same for all TM ions. Interestingly, from Table~\ref{tab:hub_param} we can see that Hubbard $U$'s for different TM elements of a given material do not change in a monotonic way upon Li intercalation. This finding is consistent with Ref.~\cite{Cococcioni:2019SM}. In other words, the $U$ values \textit{do not} change in a digital manner in contrast to the L\"owdin occupations in Fig.~3 of the main text. This fact is a consequence of the Hubbard parameters being computed self-consistently with Li content: when additional Li ions are inserted in the crystal the structural readjustment leads to a general change of the Hubbard parameters of all TM ions, although the closest to the additional Li are typically affected the most.

\section{Crystal structure parameters}
\label{sec:lattice_parameters}

\begin{figure*}[h!]
  \includegraphics[width=0.83\linewidth]{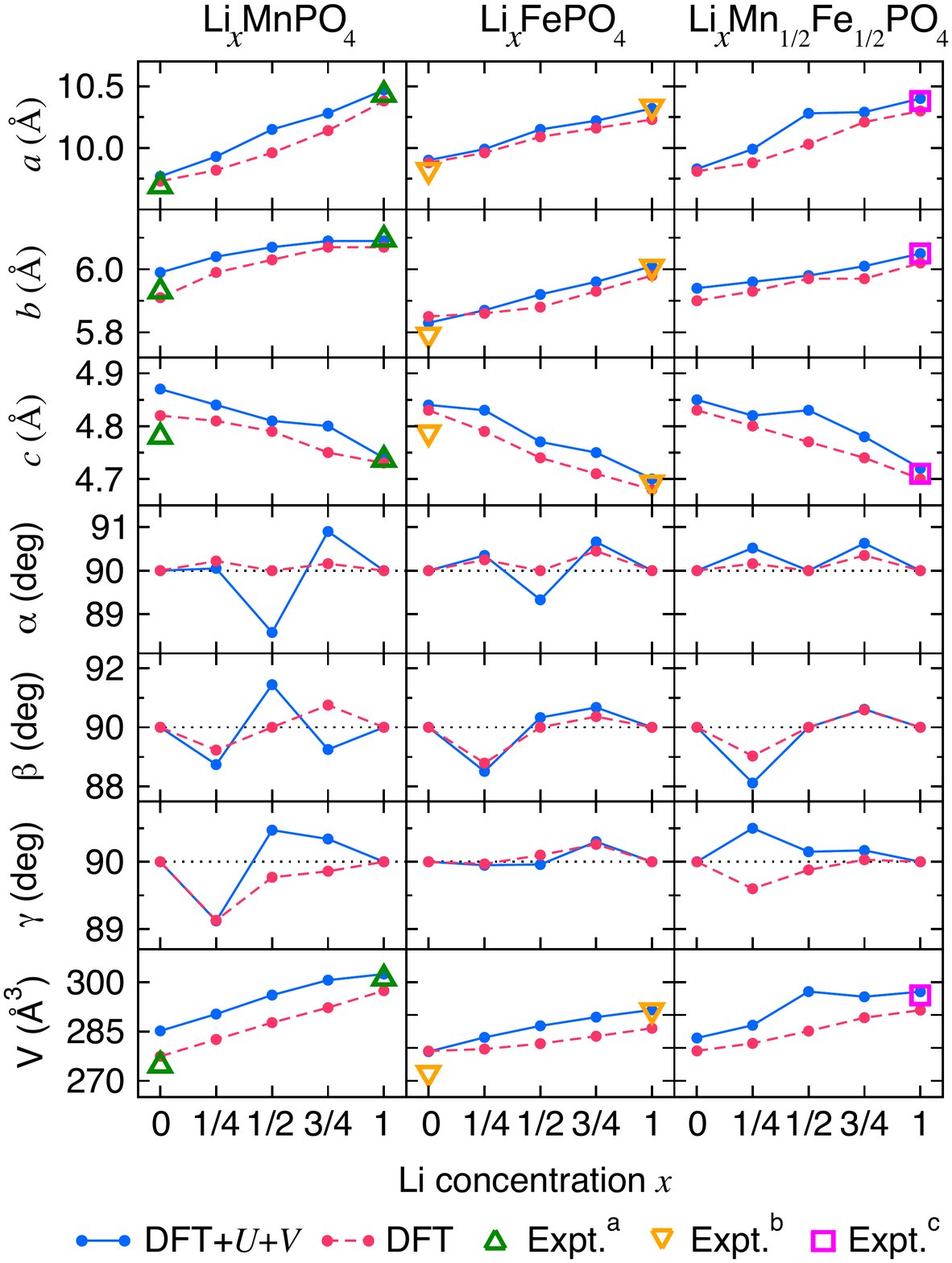}
   \caption{Lattice parameters ($a$, $b$, $c$), angles ($\alpha$, $\beta$, $\gamma$), and the cell volume ($V$) of Li$_x$MnPO$_4$, Li$_x$FePO$_4$, and Li$_x$Mn$_{1/2}$Fe$_{1/2}$PO$_4$ at $x=0,1/4,1/2,3/4,1$ computed using DFT and DFT+$U$+$V$. The experimental values are: Expt.$^\mathrm{a}$ is Ref.~\onlinecite{Nie:2010SM}, Expt.$^\mathrm{b}$ is Ref.~\onlinecite{Padhi:1997SM}, and Expt.$^\mathrm{b}$ is Ref.~\onlinecite{Muraliganth:2010SM}.}
\label{fig:lat_param}
\end{figure*}

Figure~\ref{fig:lat_param} shows the lattice parameters ($a$, $b$, $c$), angles ($\alpha$, $\beta$, $\gamma$), and the cell volume ($V$) of Li$_x$MnPO$_4$, Li$_x$FePO$_4$, and Li$_x$Mn$_{1/2}$Fe$_{1/2}$PO$_4$ at $x=0,1/4,1/2,3/4,1$ computed using DFT and DFT+$U$+$V$ in comparison with experiments~\cite{Nie:2010SM, Padhi:1997SM, Muraliganth:2010SM}. We do not present here the DFT+$U$ crystal structure parameters because, as stated in the main text, for some concentrations of Li the self-consistent evaluation of Hubbard parameters fails to converge. Moreover, HSE06 structural optimizations were not performed because of their large computational cost.

It can be seen in Fig.~\ref{fig:lat_param} that the lattice parameters $a$ and $b$ increase upon Li intercalation, while the lattice parameter $c$ decreases. This is in agreement with the experimental trends. At $x=1$ all computed lattice parameters are in remarkable agreement with the experimental values, with DFT+$U$+$V$ results being slightly more accurate than the DFT ones. At $x=0$ for Li$_x$FePO$_4$, the DFT and DFT+$U$+$V$ lattice parameters almost coincide and they somewhat overestimate the experimental ones. For Li$_x$MnPO$_4$, at $x=0$ the computed lattice parameter $a$ is very similar between and it is in good agreement with experiments, while the experimental value of $b$ is in-between the DFT and DFT+$U$+$V$ predictions. The main distinction between DFT and DFT+$U$+$V$ arises for the $c$ lattice parameter of the delithiated phase MnPO$_4$; the $c$ parameter is overestimated in DFT and even more in DFT+$U$+$V$~\cite{Cococcioni:2019SM}. This stronger departure from the experiment of DFT+$U$+$V$ compared to DFT influences the local geometry and makes some implicit incidence on the final properties. Nevertheless, the extent of corresponding relative error is less than 2\%. Similarly, for FePO$_4$ the optimized $c$ lattice parameter shows the largest deviation from the experimental one, though DFT and DFT+$U$+$V$ are now in better agreement.

The angles $\alpha$, $\beta$, and $\gamma$ are all $90^\circ$ at $x=0$ and $x=1$ because of the orthorhombic symmetry of the crystal, in agreement with experiments~\cite{Nie:2010SM, Padhi:1997SM, Muraliganth:2010SM}. However, at intermediate Li concentrations, $x=1/4$, $x=1/2$ and $x=3/4$, both DFT and DFT+$U$+$V$ predict distorted equilibrium crystal structures, with deviations from the orthorhombic symmetries that depend on the specific angle, the specific material, and the concentration of Li, and become slightly more pronounced in the presence of Mn. 

In Fig.~\ref{fig:lat_param}, we also show the cell volume at different Li concentrations. The general trend is that the addition of Li expands the lattice and the cell volume increases. At $x=1$ we find that the DFT+$U$+$V$ volumes are in remarkable agreement with the experimental ones while DFT volumes are somewhat underestimated. In contrast, at $x=0$ we observe that both DFT and DFT+$U$+$V$ volumes overestimate the experimental one in Li$_x$FePO$_4$, while in Li$_x$MnPO$_4$ the DFT volume is in good agreement with the experimental one while the DFT+$U$+$V$ volume is overestimated.

Overall we find that the accuracy of DFT+$U$+$V$ is fairly good for predicting the crystal structural parameters. For this reason we used the DFT+$U$+$V$ structural parameters also for the HSE06 calculations. 

\begin{table*}[t]
\centering
\begin{tabular}{l|l|cccccc|cccccccc}
\hline\hline
\multirow{2}{*}{Material} & \multirow{2}{*}{\parbox{0.5cm}{OS}} & \multicolumn{6}{c}{DFT}                   & \multicolumn{6}{c}{DFT+$U$+$V$}             \\ \cline{3-14}
                                              &                                            & \parbox{1cm}{$M$-O1}   & \parbox{1cm}{$M$-O2}   & \parbox{1cm}{$M$-O3}   & \parbox{1.1cm}{$M$-O3$'$} & \parbox{1.1cm}{$\langle M$-O$\rangle$} & \parbox{1cm}{$\mathcal{V}$} & \parbox{1cm}{$M$-O1}   & \parbox{1cm}{$M$-O2}   & \parbox{1cm}{$M$-O3}   & \parbox{1.1cm}{$M$-O3$'$} & \parbox{1.1cm}{$\langle M$-O$\rangle$} & \parbox{1cm}{$\mathcal{V}$}   \\ \hline\hline
LiFePO$_4$                                    & Fe$^{2+}$                                  & 2.08 & 2.18 & 2.03 & 2.23 & 2.13 & 12.12         & 2.11 & 2.20 & 2.06 & 2.25 & 2.16 & 12.56           \\      
FePO$_4$                                      & Fe$^{3+}$                                  & 1.91 & 1.92 & 2.04 & 2.13 & 2.03 & 10.59         & 1.92 & 1.93 & 2.04 & 2.15 & 2.04 & 10.79           \\ \hline
LiMnPO$_4$                                    & Mn$^{2+}$                                  & 2.12 & 2.23 & 2.10 & 2.25 & 2.18 & 12.94         & 2.14 & 2.26 & 2.13 & 2.27 & 2.20 & 13.37           \\
MnPO$_4$                                      & Mn$^{3+}$                                  & 1.85 & 1.86 & 1.96 & 2.32 & 2.05 & 10.46         & 1.88 & 1.88 & 1.98 & 2.38 & 2.08 & 10.87           \\ \hline
\multirow{2}{*}{LiMn$_{1/2}$Fe$_{1/2}$PO$_4$} & Fe$^{2+}$                                  & 2.08 & 2.17 & 2.04 & 2.24 & 2.14 & 12.21         & 2.10 & 2.20 & 2.07 & 2.26 & 2.16 & 12.64           \\
                                              & Mn$^{2+}$                                  & 2.12 & 2.34 & 2.09 & 2.24 & 2.19 & 12.82         & 2.15 & 2.26 & 2.12 & 2.26 & 2.20 & 13.28           \\ \hline
\multirow{2}{*}{Mn$_{1/2}$Fe$_{1/2}$PO$_4$}   & Fe$^{3+}$                                  & 1.89 & 1.91 & 2.06 & 2.13 & 2.03 & 10.60         & 1.90 & 1.92 & 2.08 & 2.15 & 2.05 & 10.83           \\
                                              & Mn$^{3+}$                                  & 1.85 & 1.87 & 1.95 & 2.32 & 2.04 & 10.44         & 1.89 & 1.90 & 1.97 & 2.35 & 2.07 & 10.83           \\
\hline\hline
\end{tabular}
\caption{Bond lengths $M$-O1, $M$-O2, $M$-O3, $M$-O3$'$ (in \AA), average over six bond lengths $\langle M$-O$\rangle$ (in \AA), and the volumes $\mathcal{V}$ (in \AA$^3$) of $M$O$_6$ octahedra computed using DFT and DFT+$U$+$V$, where $M=$Fe, Mn. The oxidation state (OS) of transition-metal elements $M$ is also shown.}
\label{tab:Bond_lengths}
\end{table*}

\begin{figure*}[h!]
  \includegraphics[width=0.35\linewidth]{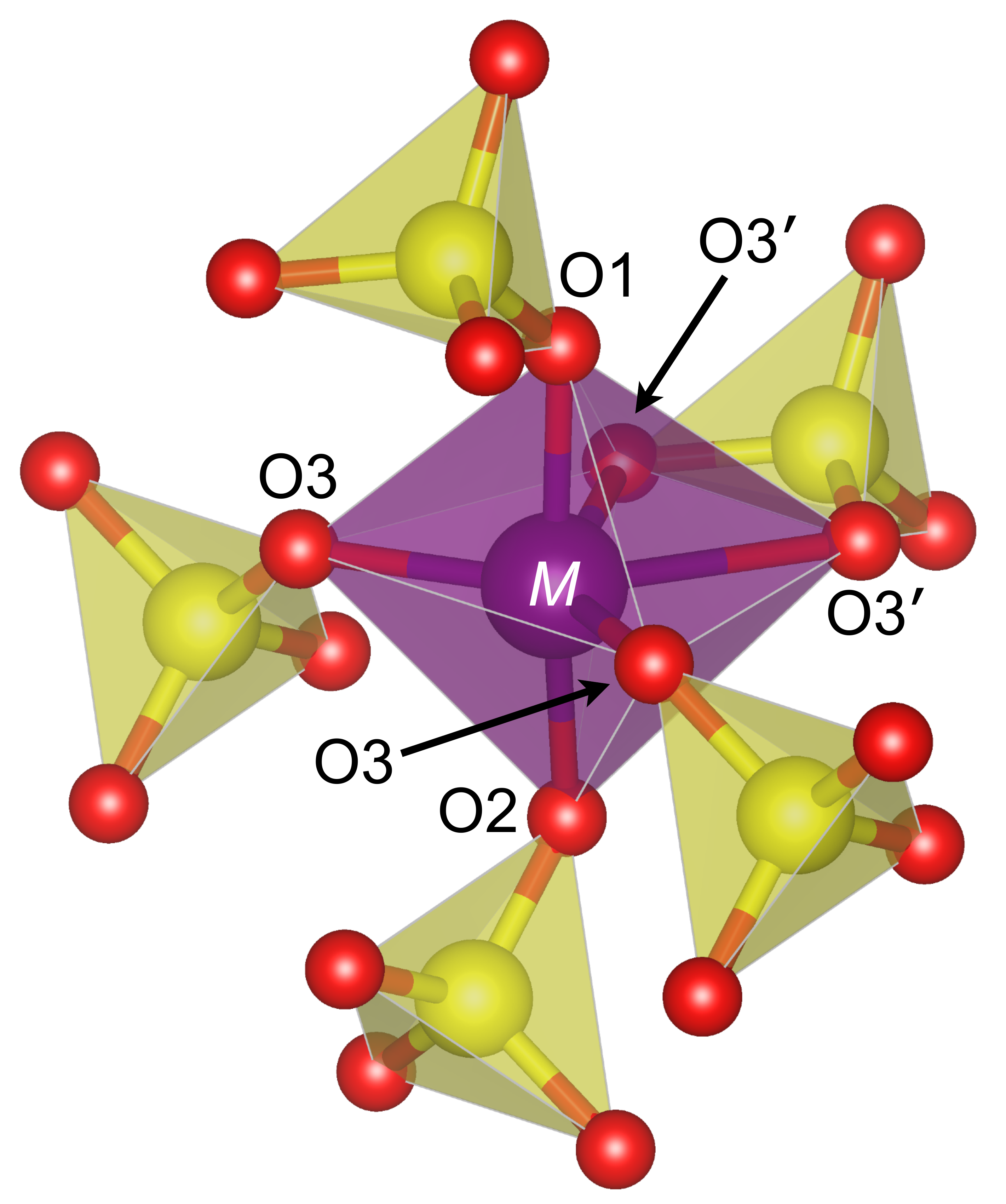}
   \caption{Local networking geometry of $M$O$_6$ octahedra ($M=$ Fe, Mn) and PO$_4$ tetrahedra in phospho-olivines. Transition-metal element $M$ is indicated in purple, P atoms in yellow, and O atoms in red. Four types of O atoms are highlighted: O1, O2, O3, and O3$'$. Rendered using \textsc{VESTA}~\cite{Momma:2008SM}.}
\label{fig:octahedra}
\end{figure*}

It is useful to comment on the structural data of the $M$O$_6$ octahedra ($M=$ Fe, Mn) upon the lithiation/delithiation. The results are shown in Table~\ref{tab:Bond_lengths}. We use the conventional notation for the four kinds of bonds~\cite{Yamada:2006SM}: $M$-O1, $M$-O2, $M$-O3, and $M$-O3$'$. The directions of $M$-O1 and $M$-O2 are axial, while two of $M$-O3 and two of $M$-O3$'$ are equatorial (see Fig.~\ref{fig:octahedra}). The $M$O$_6$ octahedra belong to $C_s$ point group symmetry, and the symmetry is conserved upon lithiation/delithiation~\cite{Asari:2011SM}. From Table~\ref{tab:Bond_lengths} we can see that all the Mn-O bonds contract when Mn atoms are oxidized (Mn$^{2+} \rightarrow $ Mn$^{3+}$), except for Mn-O3$'$ which, instead, elongate. In contrast, in the Fe-containing olivines all the Fe-O bonds contract upon the oxidation of Fe atoms (Fe$^{2+} \rightarrow $ Fe$^{3+}$). These findings are in agreement with previous DFT+$U$ studies~\cite{Seo:2010SM, Asari:2011SM, Johannes:2012SM, Piper:2013SM} and in line with experiments~\cite{Yamada:2006SM}. Both DFT and DFT+$U$+$V$ show the same trends, and they differ only quantitatively. The unique elongation of the Mn-O3$'$ bonds against the volume shrinkage of the MnO$_6$ octahedra results in a pronounced distortion of these structural units. This peculiar deformation pattern of the MnO$_6$ octahedra can be understood, along the observations made in Ref.~\cite{Yamada:2006SM}, by looking at Fig.~\ref{fig:octahedra}, where one of these octahedra is shown with neighbor structural groups. While O1, O2, and two equatorial O3 are each corner-shared with four separate PO$_4$ tetrahedra, the O3$'$-O3$'$ edge is shared with a single PO$_4$ tetrahedron. The rigidity of the PO$_4$ structural units is probably what forces an elongation of the Mn-O3$'$ bonds upon MnO$_6$ volume contraction~\cite{Yamada:2006SM}. This distortion is not a strict Jahn-Teller type, because there is no symmetry reduction in MnO$_6$ (the $C_s$ point group symmetry is preserved)~\cite{Asari:2011SM}. This type of distortion in Mn-containing phospho-olivines is called a \textit{pseudo Jahn-Teller distortion}~\cite{Piper:2013SM}.

\section{Partially delithiated structures and formation energies}
\label{sec:delithiated_structures}

\begin{figure*}[h!]
  \includegraphics[width=0.99\linewidth]{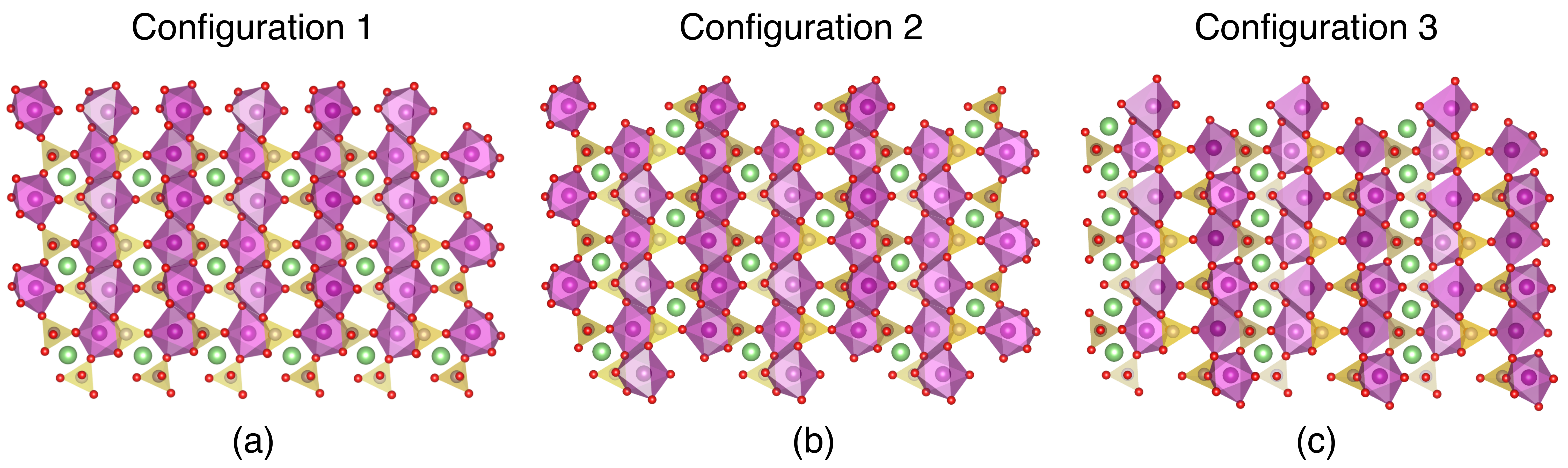}
   \caption{Three symmetrically inequivalent configurations of Li arrangements in phospho-olivines at $x=1/2$ modeled using a unit cell with four formula units (here we show $3 \times 3 \times 1$ supercells for clarity): (a)~Configuration~1, (b)~Configuration~2, and (c)~Configuration~3. Transition-metal elements $M$ are indicated in purple, P atoms in yellow, O atoms in red, and Li atoms in green. Rendered using \textsc{VESTA}~\cite{Momma:2008SM}.}
\label{fig:configurations}
\end{figure*}

In this section we analyze the stability of the three studied phospho-olivines at intermediate Li concentrations $x$. It was shown theoretically~\cite{Morgan:2004SM, OuYang:2004SM, Islam:2005bSM} and experimentally~\cite{Nishimura:2008SM} that Li diffusion is one-dimensional. Here, we consider all symmetrically inequivalent Li arrangements in the unit cell containing four formula units. All possible symmetry-distinct decorations of the four Li sites give seven structures: two end members ($x=0$ and $x=1$), one structure at each of $x=1/4$ and $x=3/4$, and three structures at $x=1/2$~\cite{Zhou:2004SM, Ong:2011SM}. At $x=1/2$, we call the three structures as ``Configuration 1'', ``Configuration 2'', and ``Configuration 3'', and these are shown in Fig.~\ref{fig:configurations}.

In Table~\ref{tab:delithiated_structures} we show the total energy differences for the three Li configurations for Li$_{1/2}$MnPO$_4$, Li$_{1/2}$FePO$_4$, and Li$_{1/2}$Mn$_{1/2}$Fe$_{1/2}$PO$_4$ computed using DFT, DFT+$U$+$V$, and HSE06. We note that we performed structural optimizations for each configuration using DFT and DFT+$U$+$V$, and for HSE06 calculations we used the DFT+$U$+$V$ geometries (as explained in Sec.~\ref{sec:technical_details}). Moreover, for each configuration we performed independent self-consistent calculations of Hubbard parameters within DFT+$U$+$V$ using DFPT. As can be seen from Table~\ref{tab:delithiated_structures}, we find that Configuration~1 is the lowest-energy structure for all materials and for all methods considered here with only one exception: for Li$_{1/2}$FePO$_4$ using HSE06 we find that Configuration~2 is the lowest-energy structure. We note that the Configuration~1 was also found to be the lowest-energy structure for Li$_{1/2}$FePO$_4$ using DFT+$U$+$V$ in Ref.~\cite{Cococcioni:2019SM}. Therefore, for the sake of consistency when comparing the accuracy of various theoretical methods, we always use Configuration~1 throughout the whole study.

\begin{table*}[h!]
\centering
\renewcommand{\arraystretch}{1.4}
\begin{tabular}{c|c|ccc}
\hline\hline
\parbox{3.5cm}{Material} & \parbox{2.3cm}{Configuration}  & \parbox{2.5cm}{DFT} & \parbox{2.5cm}{DFT+$U$+$V$} & \parbox{2.5cm}{HSE06} \\ 
\hline
                                     & 1 & 0   & 0   & 0     \\
Li$_{1/2}$MnPO$_4$                   & 2 & 5   & 17  & 12    \\
                                     & 3 & 107 & 81  & 83    \\ \hline
                                     & 1 & 0   & 0    & 153  \\
Li$_{1/2}$FePO$_4$                   & 2 & 10  & 84   & 0    \\
                                     & 3 & 101 & 237  & 39   \\ \hline
                                     & 1 & 0   & 0    & 0   \\
Li$_{1/2}$Mn$_{1/2}$Fe$_{1/2}$PO$_4$ & 2 & 8   & 7    & 10  \\
                                     & 3 & 103 & 143  & 162 \\
\hline\hline
\end{tabular}
\caption{Total energy differences (in meV/f.u.) for three different configurations of Li arrangements in Li$_{1/2}$MnPO$_4$, Li$_{1/2}$FePO$_4$, and Li$_{1/2}$Mn$_{1/2}$Fe$_{1/2}$PO$_4$ computed using DFT, DFT+$U$+$V$, and HSE06. The zero of energy in each case corresponds to the lowest-energy configuration, while positive values correspond to energies of the other two configurations relative to the lowest-energy configuration.}
\label{tab:delithiated_structures}
\end{table*}

\begin{figure*}[t]
  \includegraphics[width=0.76\linewidth]{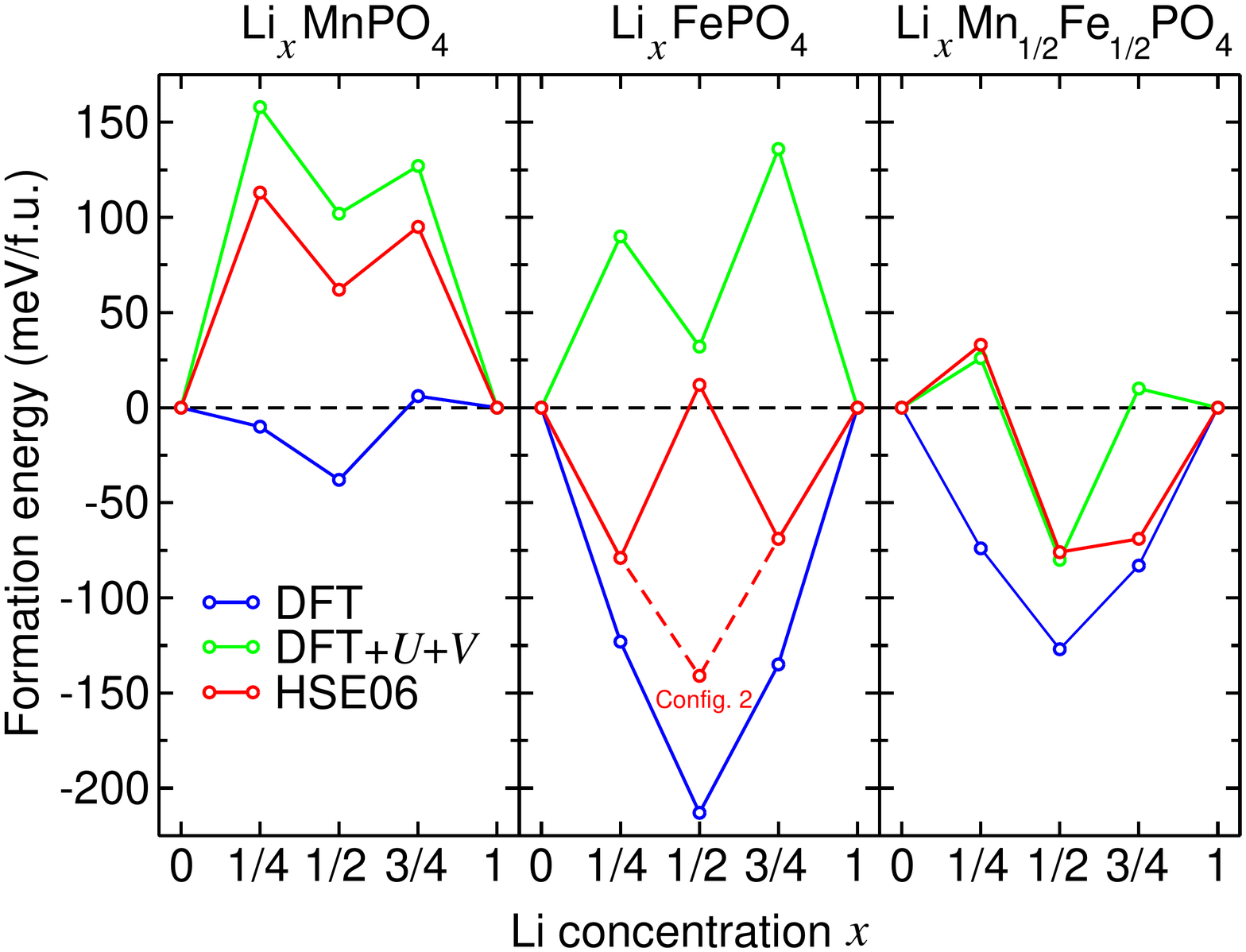}
   \caption{Formation energy (in meV/f.u.) of Li$_x$MnPO$_4$, Li$_x$FePO$_4$, and Li$_x$Mn$_{1/2}$Fe$_{1/2}$PO$_4$ computed using DFT, DFT+$U$+$V$, and HSE06. All calculations at $x=1/2$ were performed using the Configuration~1, except Li$_{1/2}$FePO$_4$ for which we also show the data point obtained using the Configuration~2 (dashed lines).}
\label{fig:formation_energies}
\end{figure*}

The formation energy $E_f$ of a compound Li$_x$S is computed as~\cite{Cococcioni:2019SM}:
\begin{equation*}
    E_f(x) = E(\mathrm{Li}_x\mathrm{S}) - x E(\mathrm{Li} \mathrm{S}) - (1 - x) E(\mathrm{S}) ,
    \label{eq:formation_energy}
\end{equation*}
where S is the short-hand notation for e.g. MnPO$_4$ in Li$_x$MnPO$_4$ and similarly for other cathode materials considered in this paper. Here, $E(\mathrm{Li}_x\mathrm{S})$, $E(\mathrm{Li} \mathrm{S})$, and $E(\mathrm{S})$ are the total energies per formula unit for the compounds Li$_x$S, LiS, and S, respectively. The results are shown in Fig.~\ref{fig:formation_energies} for the three phospho-olivines studied here. As known from previous studies~\cite{Zhou:2004SM, Ong:2011SM}, plain DFT predicts negative or near-zero formation energies which contradicts the lack of experimental observation of stable compounds at intermediate Li concentrations ($0 < x < 1$) for both Li$_x$FePO$_4$ and Li$_x$MnPO$_4$ that are characterized by a two-phase reaction upon (de-)lithiation and a single voltage plateau~\cite{Padhi:1997SM, Yamada:2001SM}. We note that our DFT formation energies somewhat differ from those of Ref.~\cite{Ong:2011SM}, most likely because we use the PBEsol functional while in Ref.~\cite{Ong:2011SM} the authors used PBE. Remarkably, DFT+$U$+$V$ correctly predicts positive-valued formation energies and thus unstable Li$_x$FePO$_4$ and Li$_x$MnPO$_4$ at fractional $x$, which is in line with previous DFT+$U$ studies~\cite{Ong:2011SM}. Our HSE06 formation energies for Li$_x$FePO$_4$ agree well qualitatively with previous HSE06 studies~\cite{Ong:2011SM}: they predict stable structures at $x=1/4$ and $x=3/4$, thus contradicting experiments. It is worth mentioning that for Li$_{1/2}$FePO$_4$ the lowest-energy structure when using HSE06 is Configuration~2, which is shown as the minimum of the dashed line in the central panel of Fig.~\ref{fig:formation_energies}, and highlights a similar trend to the DFT. Therefore, for Li$_x$FePO$_4$ and Li$_x$MnPO$_4$ we find that DFT+$U$+$V$ correctly predicts the two-phase reaction in agreement with experiments, while HSE06 fails for Li$_x$FePO$_4$ as was also found in a previous study~\cite{Ong:2011SM}.

The case of Li$_x$Mn$_{1/2}$Fe$_{1/2}$PO$_4$ is of particular interest. Experimentally it was found that a two-phase reaction occurs in the range $0 \leq x < 1/2$, while for $1/2 < x \lesssim 1$ the system undergoes a single-phase reaction~\cite{Yamada:2001SM}. For $x=1/2$, i.e. at the border of the two regions,  there is no clear consensus from experiments although previous DFT+$U$ studies show that this composition lies on the convex hull thus resulting stable~\cite{Loftager:2019SM}. At $x=1/4$, the formation energies from DFT+$U$+$V$ and HSE06 agree remarkably well and both are positive meaning that the structure is unstable in agreement with the experimentally observed two-phase reaction. At $x=1/2$, the formation energies from DFT+$U$+$V$ and HSE06 again agree very well and are both negative meaning that the structure is stable in agreement with Ref.~\cite{Loftager:2019SM}. However, at $x=3/4$ the formation energies from DFT+$U$+$V$ and HSE06 disagree: the former gives a positive value while the latter gives a negative value. In other words, at $x=3/4$, HSE06 predicts a stable phase (since it lies on the convex hull) whose existence agrees with the experimentally observed single-phase reaction, while DFT+$U$+$V$ predicts this composition to be above the convex hull, thus contradicting experiments. We note that previous DFT+$U$ studies~\cite{Loftager:2019SM} also predicted the Li$_{3/4}$Mn$_{1/2}$Fe$_{1/2}$PO$_4$ structure to be unstable. It is worth noting that the DFT+$U$+$V$ formation energy at $x=3/4$ is very small (10~meV/f.u.) and that it is sensitive to various details of the calculations. In particular, we remind that here we consider only one specific case of the arrangement of the Mn and Fe atoms in Li$_x$Mn$_{1/2}$Fe$_{1/2}$PO$_4$ and only one specific antiferromagnetic ordering. A more thorough study of this structure would require considering all possible arrangements of TM atoms and spins to identify the lowest-energy one. In addition, we recall that here we have neglected the effects from the configuration entropy, the inclusion of which might also influence the trends for the formation (free) energies. Hence, we argue that the findings presented for the formation energies of Li$_x$Mn$_{1/2}$Fe$_{1/2}$PO$_4$ are possibly not conclusive and that a more extensive study is needed to clarify this point.

\section{Magnetic moments}
\label{sec:magnetic_moments}

\begin{figure*}[h!]
  \includegraphics[width=0.9\linewidth]{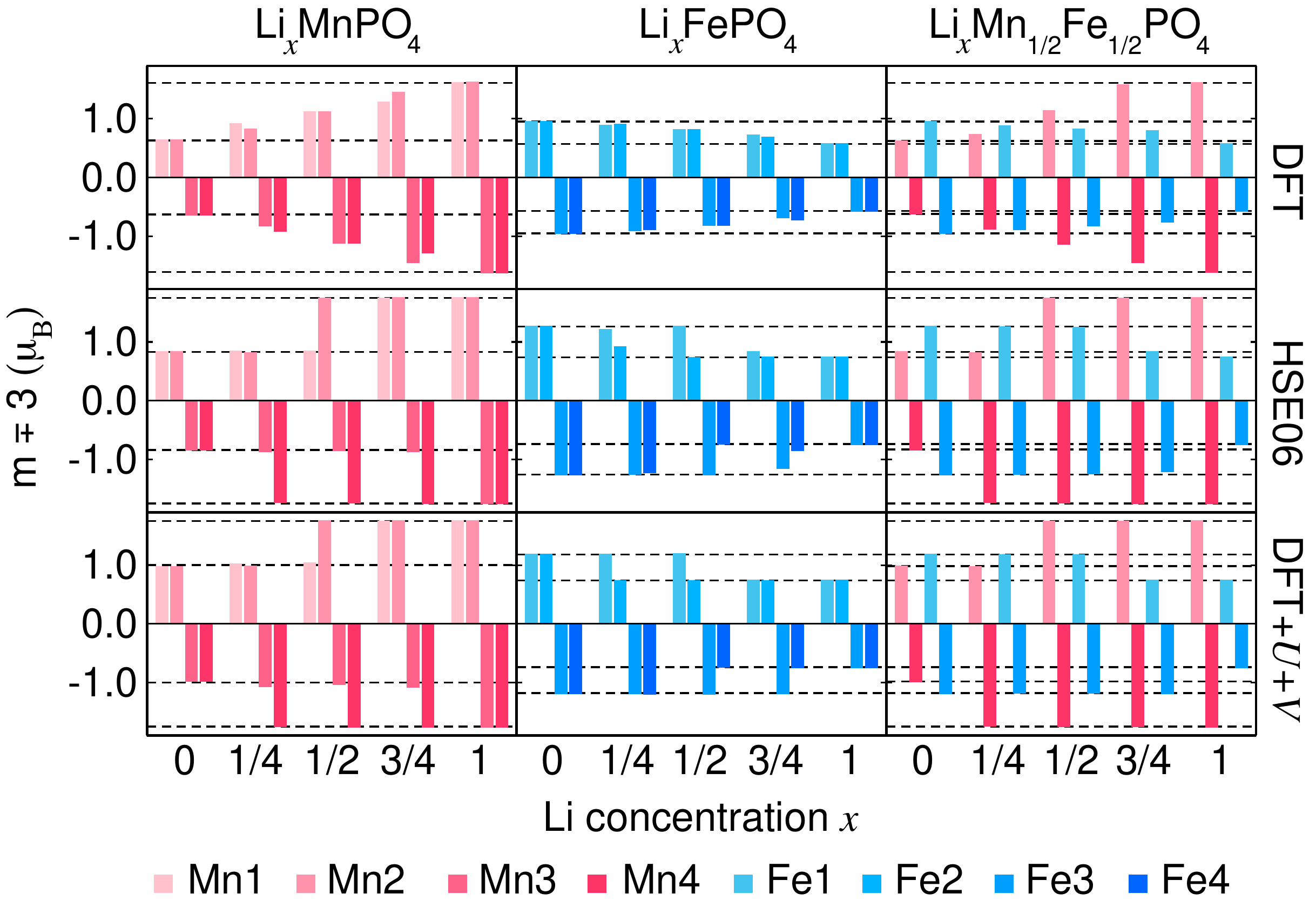}
   \caption{Magnetic moments (shifted by $\mp 3~\mu_\mathrm{B}$ for the sake of clearer comparison) for Mn and Fe atoms in Li$_x$MnPO$_4$, Li$_x$FePO$_4$, and Li$_x$Mn$_{1/2}$Fe$_{1/2}$PO$_4$ at $x=0, 1/4, 1/2, 3/4, 1$ computed using three approaches (DFT, DFT+$U$+$V$, and HSE06). The horizontal dashed lines correspond to the magnetic moments of the end elements ($x=0$ and $x=1$). For each material there are four transition-metal atoms, each of which is represented with a bar.}
\label{fig:magn_moments}
\end{figure*}

Figure~\ref{fig:magn_moments} shows the relative magnetic moments for Mn and Fe atoms in Li$_x$MnPO$_4$, Li$_x$FePO$_4$, and Li$_x$Mn$_{1/2}$Fe$_{1/2}$PO$_4$ at $x=0, 1/4, 1/2, 3/4, 1$ computed using three approaches (DFT, DFT+$U$+$V$, and HSE06). The exact values of the magnetic moments at $x=0$ and $x=1$ are reported in Table~I of the main text and in Table~\ref{tab:OS_MnFe}. These magnetic moments were computed via the projection method, i.e. by computing the atomic occupation matrix [see Eq.~5 in the main text], then by diagonalizing it, and eventually by computing $m = \sum_{i=1}^5 (\lambda_i^\uparrow - \lambda_i^\downarrow)$ using the data in Table~\ref{tab:OS_MnFe}. 

Before we proceed to the detailed analysis of Fig.~\ref{fig:magn_moments}, it is useful to compare the computed magnetic moments with the available experimental values. On the one hand, for Li$_x$MnPO$_4$, at $x=0$ the experimental magnetic moments are not known (to the best of our knowledge), while at $x=1$ they are 4.28~\cite{Gnewuch:2020SM} and 5.20~$\mu_\mathrm{B}$~\cite{Newnham:1965SM}. These experimental values are largely scattered which complicates a comparison with our theoretical predictions. Nevertheless, from Table~I in the main text we can see that both DFT+$U$+$V$ and HSE06 predict the magnetic moment to be 4.75~$\mu_\mathrm{B}$ which falls right between the two experimental values. The predictions by DFT and DFT+$U$ are 4.62 and 4.76~$\mu_\mathrm{B}$, respectively, which are also intermediate to the experimental values. On the other hand, for Li$_x$FePO$_4$, at $x=0$ the experimental magnetic moment is 4.15~$\mu_\mathrm{B}$ while at $x=1$ it is 4.19~$\mu_\mathrm{B}$~\cite{Rousse:2003SM}. At $x=0$, DFT+$U$+$V$ and HSE06 predict the magnetic moments to be 4.18 and 4.26~$\mu_\mathrm{B}$, respectively, and at $x=1$ the magnetic moments are 3.74 from both methods. DFT and DFT+$U$ provide quite similar values (see Table~I in the main text). Therefore, at $x=0$, the DFT+$U$+$V$ magnetic moment falls within the experimental range and proves more accurate than HSE06 and other methods considered here. However, at $x=1$ the computed magnetic moments are smaller than the experimental ones. It would be important to have novel and more accurate experiments on Li$_x$FePO$_4$ to verify whether the trend and values of Ref.~\cite{Rousse:2003SM} are correct and, in particular, whether and how Fe$^{3+}$ and Fe$^{2+}$ can assume very similar magnetic moments. Finally, we are not aware of experimental measurements of the magnetic moments in Li$_x$Mn$_{1/2}$Fe$_{1/2}$PO$_4$.

Now let us discuss Fig.~\ref{fig:magn_moments}. Our main goal here (as in the main text for L\"owdin occupations) is to compare the accuracy of the DFT+$U$+$V$ approach versus HSE06 for predicting a relative change in magnetic moments upon (de-)lithiation of phospho-olivines. In the case of Li$_x$MnPO$_4$, we can see that DFT+$U$+$V$ and HSE06 agree remarkably well and both show a digital change in the magnetic moments: adding one Li$^+$ ion and one electron to the cathode during the lithiation process leads to the change in the magnetic moment of only one Mn ion (that accepts this extra electron) while all other Mn ions remain unchanged. This process continues when we go on with the Li intercalation until eventually all Mn ions switch their magnetic moments between the values characteristic of their +3 and +2 OS. Thus, these two approaches successfully describe the mixed-valence nature of the Li$_x$MnPO$_4$ compound that contains two types of Mn ions, Mn$^{3+}$ and Mn$^{2+}$, at $x=1/4, 1/2, 3/4$. In contrast, DFT fails to localize an extra electron on one of the Mn ions; as a consequence, also magnetization has a gradual collective change on all the Mn ions, reflecting the fractional OS they assume. In the case of Li$_x$FePO$_4$, we find similar trends with the difference that here only DFT+$U$+$V$ shows a digital change in the magnetic moments while HSE06 features a more gradual change in the magnetic moments of Fe atoms. In full consistency with the previous observations, in Li$_x$Mn$_{1/2}$Fe$_{1/2}$PO$_4$ we find that both DFT+$U$+$V$ and HSE06 predict a digital change of the magnetic moments of Mn ions, while for Fe ions again we find that only DFT+$U$+$V$ achieves a digital switch of the magnetic moments.

\begin{figure*}[h!]
  \subfigure[]{\includegraphics[width=0.47\linewidth]{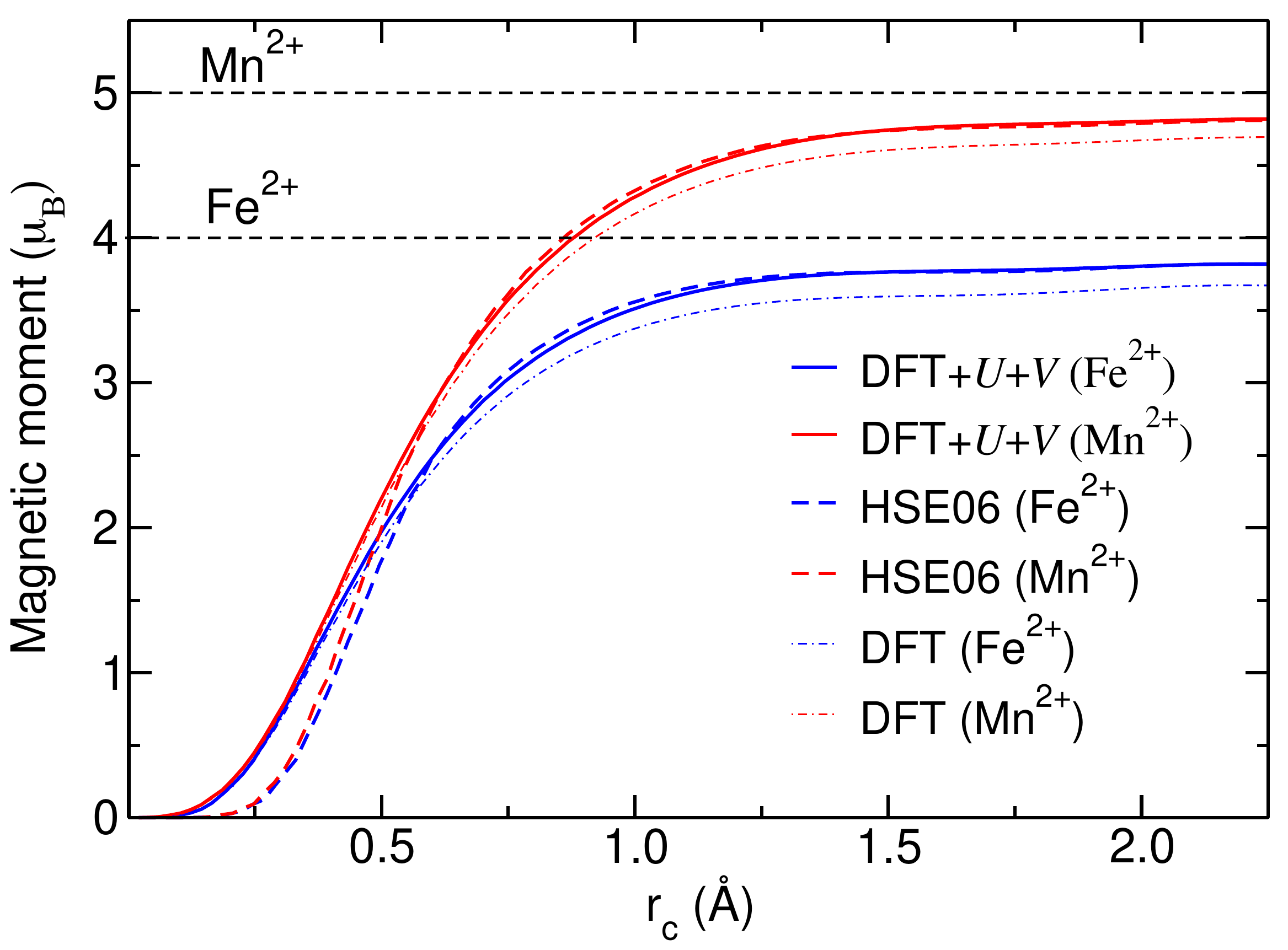}}
  \hspace{0.3cm}
  \subfigure[]{\includegraphics[width=0.47\linewidth]{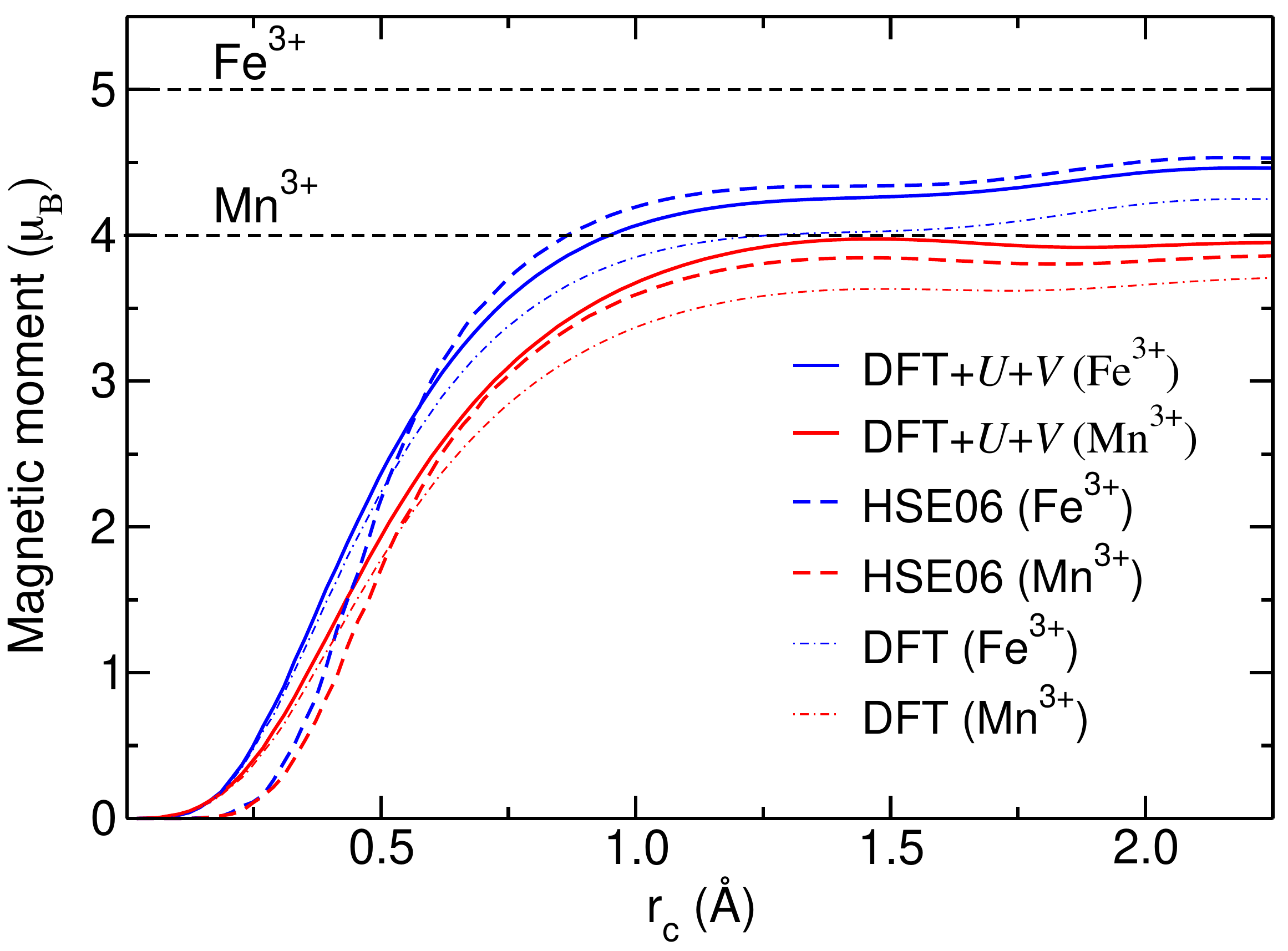}}
   \caption{Magnetic moments for Mn and Fe atoms as a function of the cutoff radius $\mathrm{r}_\mathrm{c}$ in Li$_x$Mn$_{1/2}$Fe$_{1/2}$PO$_4$ at (a)~$x=1$ and (b)~$x=0$, computed using three approaches (DFT, DFT+$U$+$V$, and HSE06). The horizontal dashed lines correspond to the nominal oxidation state.}
\label{fig:magn_moments_2}
\end{figure*}

Finally, the magnetic moments can be useful for the determination of the OS of transition-metal ions. Following Ref.~\cite{Reed:2002SM} we compute the magnetic moments by integrating the difference between the spin-up and spin-down components of the spin-charge density inside a sphere around an ion. Since the magnetic moment is very sensitive to the cutoff radius $\mathrm{r}_\mathrm{c}$ of the sphere, we compute it for many values of $\mathrm{r}_\mathrm{c}$~\cite{Reed:2002SM} and study its dependence on this parameter. The result is shown in Fig.~\ref{fig:magn_moments_2} for the case of Li$_x$Mn$_{1/2}$Fe$_{1/2}$PO$_4$ at $x=1$ and $x=0$. We can see that the magnetic moment increases steeply as we perform the integration around the transition-metal ion, but quickly reaches a plateau in the interstitial region (O atoms do not contribute significantly to the magnetization). For $x=1$, the plateau values for Mn and Fe are very close to the nominal magnetic moments of 5.0 and 4.0~$\mu_\mathrm{B}$, respectively, which confirms that both transition-metal elements are in the $+2$ OS. Interestingly, the plateau values within DFT+$U$+$V$ and HSE06 are on top of each other, while DFT has smaller values. Moreover, it is useful to mention that the magnetic moments within DFT+$U$+$V$ and HSE06 increase with different rates in the range of $\mathrm{r}_\mathrm{c}$ from 0 to $0.5$~\AA, which shows that there are different levels of localization of the density close to the nucleus within these two approaches. For $x=0$ the Mn's magnetic moment plateau value is very close to the nominal value of 4.0~$\mu_\mathrm{B}$ (thus confirming the +3 OS of the ion). At $x=0$, for Fe the magnetic moment shows a  residual dependence on the cut-off radius (determining a slightly ``wavy'' behavior) all the way to a distance of 2.25~\AA\ from the nucleaus. Most importantly, Fe's magnetic moment converges to about 4.5~$\mu_\mathrm{B}$ which makes the association with the +3 OS less obvious. The differences between the magnetic moments  predicted by various approaches in Fig.~\ref{fig:magn_moments_2}~(b) are larger than in Fig.~\ref{fig:magn_moments_2}~(a). Therefore, we can conclude that while the method of Ref.~\cite{Reed:2002SM} works well for the determination of the OS of transition-metal ions in LiMn$_{1/2}$Fe$_{1/2}$PO$_4$, it is less transparent and unambiguous in the case of Mn$_{1/2}$Fe$_{1/2}$PO$_4$. For this reason in the manuscript we used the projection-based method of Ref.~\cite{Sit:2011SM} for the determination of the OS; this in fact worked well in all cases considered in this work.

\section{Population analysis for Li$_x$Mn$_{1/2}$Fe$_{1/2}$PO$_4$}
\label{sec:OS}

Table~\ref{tab:OS_MnFe} presents the population analysis data for the $3d$ shell of Mn and Fe atoms in Li$_x$Mn$_{1/2}$Fe$_{1/2}$PO$_4$ at $x=0$ and $x=1$ computed using four approaches (DFT, DFT+$U$, DFT+$U$+$V$, and HSE06) and the nominal data. More specifically, it shows the eigenvalues of the site-diagonal ($I=J$) atomic occupation matrix $n^{I\sigma}_{mm'}$ of size $5 \times 5$ [see Eq.~(5) in the main text] in the spin-up ($\sigma = \uparrow$ : $\lambda_i^\uparrow$) and spin-down ($\sigma = \downarrow$ : $\lambda_i^\downarrow$) channels, L\"owdin occupations $n = \sum_{i=1}^5 (\lambda_i^\uparrow + \lambda_i^\downarrow)$, magnetic moments $m = \sum_{i=1}^5 (\lambda_i^\uparrow - \lambda_i^\downarrow)$, and the OS determined using the method of Ref.~\cite{Sit:2011SM}.

By comparing the data in Table~\ref{tab:OS_MnFe} with the one in Table~I in the main text we can see that the trends are the same. In other words, in the mixed olivine Li$_x$Mn$_{1/2}$Fe$_{1/2}$PO$_4$, the occupations $n$, magnetic moments $m$, and eigenvalues $\lambda_i^\sigma$ are essentially the same as in corresponding olivines Li$_x$FePO$_4$ and Li$_x$MnPO$_4$. This means that mixing of Mn and Fe atoms in the same olivine compound does not lead to changes in their "pristine" electrochemical properties. Therefore, the discussion and the analysis presented in Sec.~III.A applies also to Li$_x$Mn$_{1/2}$Fe$_{1/2}$PO$_4$ in exactly the same way.

By using the method of Ref.~\cite{Sit:2011SM}, the analysis of the eigenvalues of the atomic occupation matrix in Table~\ref{tab:OS_MnFe} reveals that at $x=0$ both Mn and Fe are in the $+3$ OS, while at $x=1$ both of them are in the $+2$ OS. By performing the same analysis at intermediate Li concentrations we find that for $0<x<1/2$ it is the Mn ions that change their OS from $+3$ to $+2$ while all Fe ions remain in the $+3$ OS. On the contrary, for $1/2<x<1$ it is the Fe ions that get reduced from $+3$ to $+2$ while Mn ions remain in the $+2$ OS (see Fig.~3 in the main text). These results are fully consistent with the fact that Mn$^{3+/2+}$ redox couple presents a higher reduction potential than the Fe$^{3+/2+}$ redox couple.

\begin{table*}[t]
\centering
\begin{tabular}{l|clcccccccccccccc}
\hline\hline
\parbox{1.5cm}{Method} & \parbox{1cm}{$x$} & Element & $\lambda_1^{\uparrow}$ & $\lambda_2^{\uparrow}$ & $\lambda_3^{\uparrow}$ & $\lambda_4^{\uparrow}$ & $\lambda_5^{\uparrow}$ & \phantom{a} & $\lambda_1^{\downarrow}$ & $\lambda_2^{\downarrow}$ & $\lambda_3^{\downarrow}$ & $\lambda_4^{\downarrow}$ & $\lambda_5^{\downarrow}$ & \parbox{1cm}{$n$} & \parbox{1.15cm}{$m$ ($\mu_\mathrm{B}$)} & \parbox{1cm}{OS} \\
\hline\hline
\multirow{4}{*}{DFT}         & \multirow{2}{*}{0} & Mn & 0.43       & {\bf 0.98} & {\bf 0.98} & {\bf 0.99} & {\bf 1.00} & & 0.09 & 0.11 & 0.13 & 0.16 & 0.27       & 5.12 & 3.62 & +3 \\
                             &                    & Fe & {\bf 0.97} & {\bf 0.98} & {\bf 0.99} & {\bf 1.00} & {\bf 1.00} & & 0.15 & 0.17 & 0.17 & 0.25 & 0.26       & 5.93 & 3.95 & +3 \\ \cline{2-17}
                             & \multirow{2}{*}{1} & Mn & {\bf 0.98} & {\bf 0.99} & {\bf 0.99} & {\bf 0.99} & {\bf 1.00} & & 0.03 & 0.04 & 0.05 & 0.11 & 0.11       & 5.28 & 4.61 & +2 \\ 
                             &                    & Fe & {\bf 0.99} & {\bf 0.99} & {\bf 0.99} & {\bf 0.99} & {\bf 1.00} & & 0.06 & 0.07 & 0.13 & 0.14 & {\bf 0.98} & 6.32 & 3.57 & +2 \\ \hline
\multirow{4}{*}{DFT+$U$}     & \multirow{2}{*}{0} & Mn & 0.54       & {\bf 0.99} & {\bf 0.99} & {\bf 1.00} & {\bf 1.00} & & 0.04 & 0.05 & 0.06 & 0.09 & 0.19       & 4.95 & 4.10 & +3 \\
                             &                    & Fe & {\bf 0.99} & {\bf 0.99} & {\bf 1.00} & {\bf 1.00} & {\bf 1.00} & & 0.08 & 0.11 & 0.11 & 0.21 & 0.24       & 5.72 & 4.22 & +3 \\ \cline{2-17}
                             & \multirow{2}{*}{1} & Mn & {\bf 0.99} & {\bf 0.99} & {\bf 1.00} & {\bf 1.00} & {\bf 1.00} & & 0.02 & 0.02 & 0.03 & 0.07 & 0.08       & 5.19 & 4.76 & +2 \\ 
                             &                    & Fe & {\bf 0.99} & {\bf 0.99} & {\bf 1.00} & {\bf 1.00} & {\bf 1.00} & & 0.03 & 0.03 & 0.08 & 0.09 & {\bf 1.00} & 6.20 & 3.76 & +2 \\ \hline
\multirow{4}{*}{DFT+$U$+$V$} & \multirow{2}{*}{0} & Mn & 0.50       & {\bf 0.99} & {\bf 0.99} & {\bf 1.00} & {\bf 1.00} & & 0.05 & 0.06 & 0.08 & 0.10 & 0.22       & 4.98 & 3.98 & +3 \\
                             &                    & Fe & {\bf 0.99} & {\bf 0.99} & {\bf 1.00} & {\bf 1.00} & {\bf 1.00} & & 0.09 & 0.13 & 0.13 & 0.21 & 0.24       & 5.76 & 4.18 & +3 \\ \cline{2-17}
                             & \multirow{2}{*}{1} & Mn & {\bf 0.99} & {\bf 0.99} & {\bf 1.00} & {\bf 1.00} & {\bf 1.00} & & 0.02 & 0.02 & 0.03 & 0.08 & 0.08       & 5.21 & 4.75 & +2 \\ 
                             &                    & Fe & {\bf 0.99} & {\bf 0.99} & {\bf 1.00} & {\bf 1.00} & {\bf 1.00} & & 0.03 & 0.04 & 0.09 & 0.10 & {\bf 0.99} & 6.22 & 3.74 & +2 \\ \hline
\multirow{4}{*}{HSE06}       & \multirow{2}{*}{0} & Mn & 0.40       & {\bf 0.99} & {\bf 0.99} & {\bf 0.99} & {\bf 0.99} & & 0.06 & 0.07 & 0.08 & 0.11 & 0.23       & 4.91 & 3.83 & +3 \\
                             &                    & Fe & {\bf 0.99} & {\bf 0.99} & {\bf 0.99} & {\bf 0.99} & {\bf 1.00} & & 0.08 & 0.11 & 0.11 & 0.19 & 0.22       & 5.67 & 4.26 & +3 \\ \cline{2-17}
                             & \multirow{2}{*}{1} & Mn & {\bf 0.99} & {\bf 0.99} & {\bf 1.00} & {\bf 1.00} & {\bf 1.00} & & 0.02 & 0.02 & 0.03 & 0.08 & 0.08       & 5.21 & 4.75 & +2 \\ 
                             &                    & Fe & {\bf 0.99} & {\bf 0.99} & {\bf 1.00} & {\bf 1.00} & {\bf 1.00} & & 0.03 & 0.04 & 0.09 & 0.09 & {\bf 0.99} & 6.22 & 3.74 & +2 \\ \hline
\multirow{4}{*}{Nominal}     & \multirow{2}{*}{0} & Mn & 0.00       & {\bf 1.00} & {\bf 1.00} & {\bf 1.00} & {\bf 1.00} & & 0.00 & 0.00 & 0.00 & 0.00 & 0.00       & 4.00 & 4.00 & +3 \\
                             &                    & Fe & {\bf 1.00} & {\bf 1.00} & {\bf 1.00} & {\bf 1.00} & {\bf 1.00} & & 0.00 & 0.00 & 0.00 & 0.00 & 0.00       & 5.00 & 5.00 & +3 \\ \cline{2-17}
                             & \multirow{2}{*}{1} & Mn & {\bf 1.00} & {\bf 1.00} & {\bf 1.00} & {\bf 1.00} & {\bf 1.00} & & 0.00 & 0.00 & 0.00 & 0.00 & 0.00       & 5.00 & 5.00 & +2 \\
                             &                    & Fe & {\bf 1.00} & {\bf 1.00} & {\bf 1.00} & {\bf 1.00} & {\bf 1.00} & & 0.00 & 0.00 & 0.00 & 0.00 & {\bf 1.00} & 6.00 & 4.00 & +2 \\
\hline\hline
\end{tabular}%
\caption{Population analysis data for the $3d$ shell of Mn and Fe atoms in Li$_x$Mn$_{1/2}$Fe$_{1/2}$PO$_4$ at $x=0$ and $x=1$ computed using four approaches (DFT, DFT+$U$, DFT+$U$+$V$, and HSE06) and the nominal data. This table shows the eigenvalues of the site-diagonal occupation matrix for the spin-up ($\lambda_i^\uparrow$, $i=\overline{1,5}$) and spin-down ($\lambda_i^\downarrow$, $i=\overline{1,5}$) channels, L\"owdin occupations $n = \sum_i (\lambda_i^\uparrow + \lambda_i^\downarrow)$, magnetic moments $m = \sum_i (\lambda_i^\uparrow - \lambda_i^\downarrow)$, and the oxidation state (OS). For the sake of simplicity we dropped the atomic site index $I$ from all quantities reported here. The eigenvalues are written in the ascending order (from left to right) for each spin channel. The eigenvalues written in bold are considered as being such that correspond to fully occupied states and thus are taken into account when determining the OS according to Ref.~\cite{Sit:2011SM}.}
\label{tab:OS_MnFe}
\end{table*}

\section{Projected density of states for Li$_x$MnPO$_4$ and Li$_x$FePO$_4$}
\label{sec:PDOS}

\begin{figure*}[t]
  \includegraphics[width=0.8\linewidth]{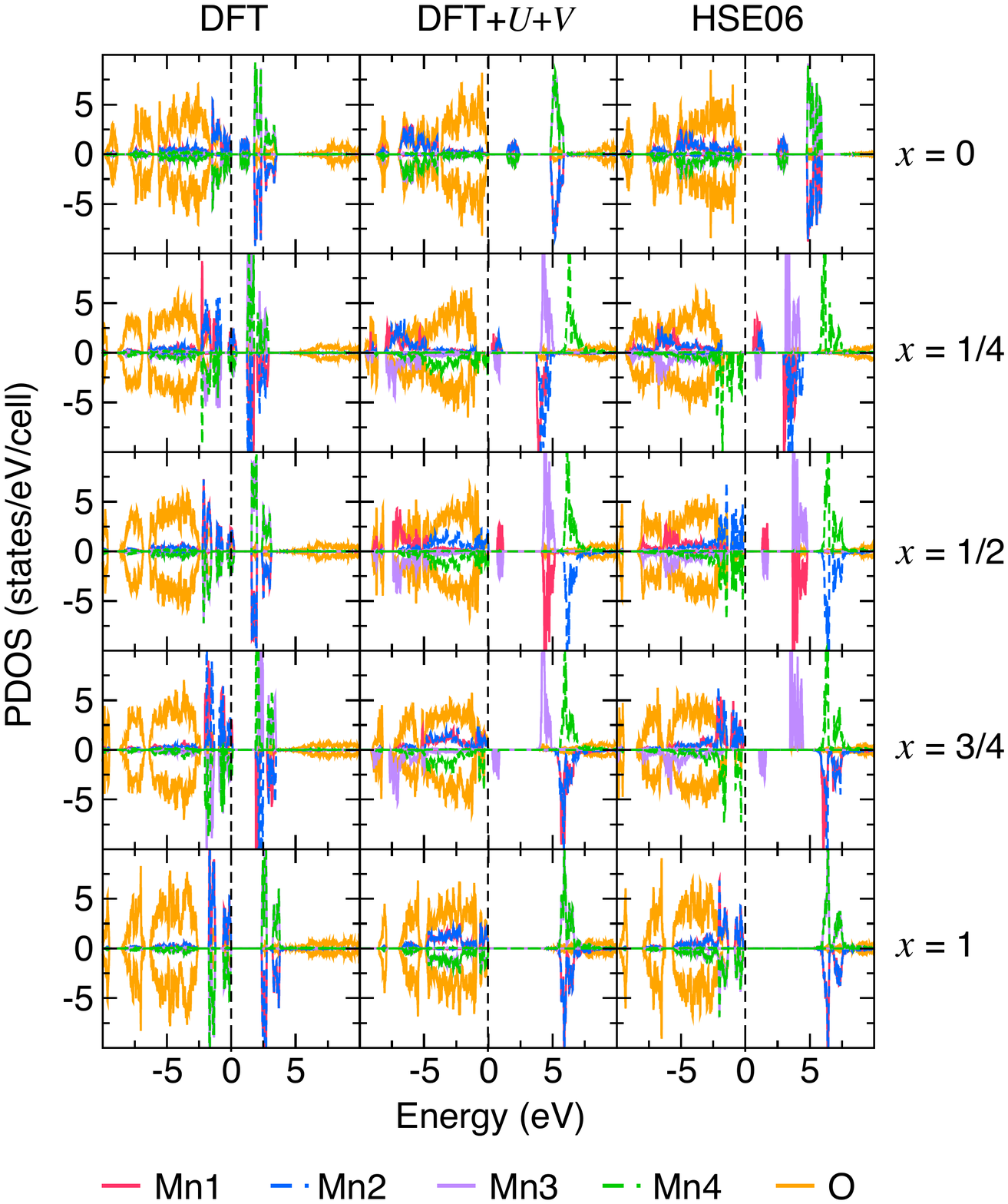}
   \caption{Spin-resolved PDOS in Li$_x$MnPO$_4$ at different concentrations of Li ($x = 0, 1/4, 1/2, 3/4, 1$) for $3d$ states of Mn1, Mn2, Mn3, Mn4 and for $2p$ states of O, computed using DFT, DFT+$U$+$V$, and HSE06. The PDOS for O-$2p$ states was obtained by summing up contributions from all O atoms in the simulation cell and it was multiplied by a factor of $1/2$ in order to have clearer comparison with the PDOS of Mn atoms. The zero of energy corresponds to the top of the valence bands in the case of insulating ground states or the Fermi level in the case of metallic ground states. The upper part of each panel corresponds to the spin-up channel, and the lower part corresponds to the spin-down channel.}
\label{fig:PDOS_LMPO}
\end{figure*}

\begin{figure*}[t]
  \includegraphics[width=0.8\linewidth]{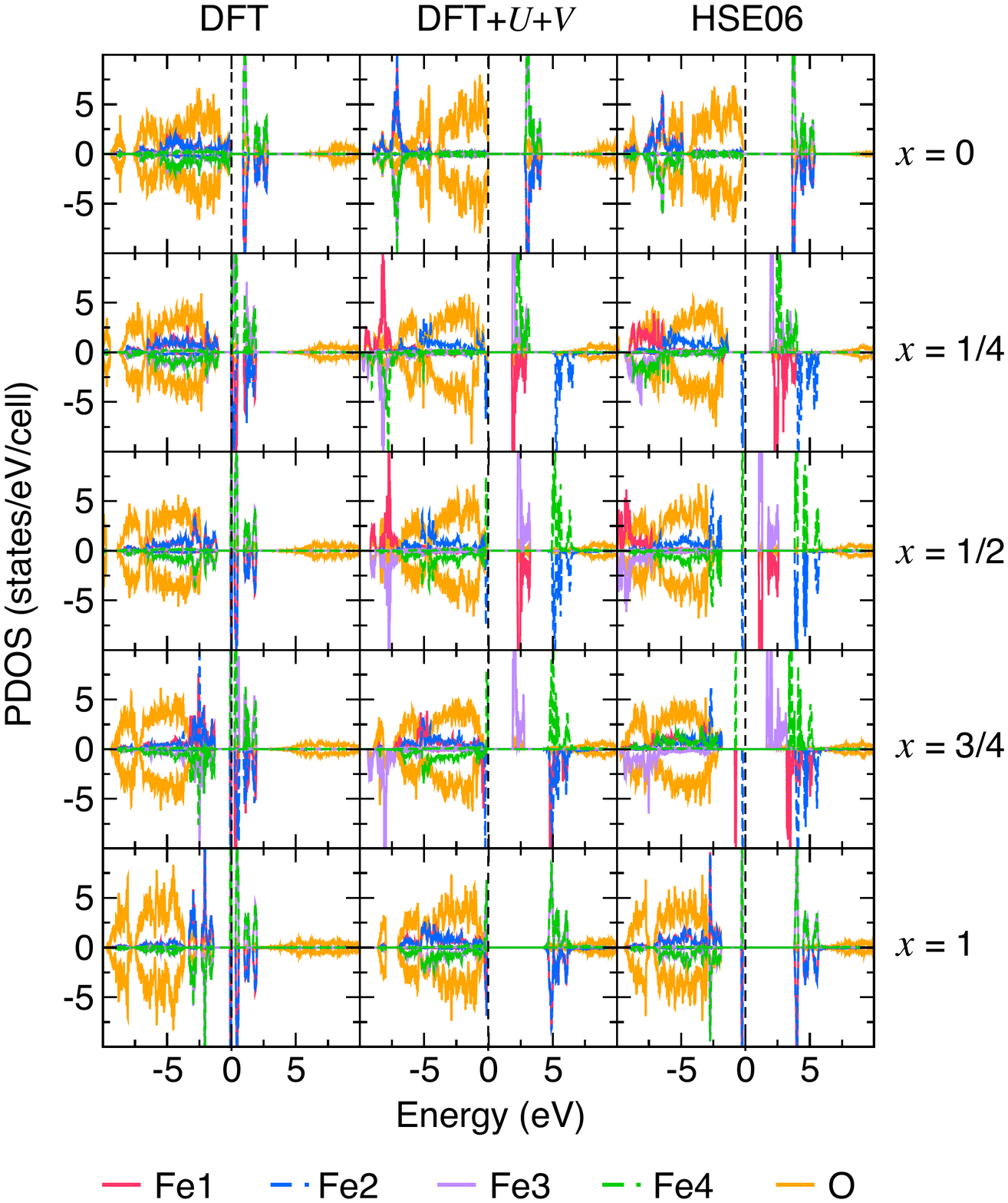}
   \caption{Spin-resolved PDOS in Li$_x$FePO$_4$ at different concentrations of Li ($x = 0, 1/4, 1/2, 3/4, 1$) for $3d$ states of Fe1, Fe2, Fe3, Fe4 and for $2p$ states of O, computed using DFT, DFT+$U$+$V$, and HSE06. The PDOS for O-$2p$ states was obtained by summing up contributions from all O atoms in the simulation cell and it was multiplied by a factor of $1/2$ in order to have clearer comparison with the PDOS of Fe atoms. The zero of energy corresponds to the top of the valence bands in the case of insulating ground states or the Fermi level in the case of metallic ground states. The upper part of each panel corresponds to the spin-up channel, and the lower part corresponds to the spin-down channel.}
\label{fig:PDOS_LFPO}
\end{figure*}

Figures~\ref{fig:PDOS_LMPO} and \ref{fig:PDOS_LFPO} show the spin-resolved PDOS for Li$_x$MnPO$_4$ and Li$_x$FePO$_4$ at different concentrations of Li ($x = 0, 1/4, 1/2, 3/4, 1$) using three approaches (DFT, DFT+$U$+$V$, and HSE06). The observations are similar to those presented for Li$_x$Mn$_{1/2}$Fe$_{1/2}$PO$_4$ in Sec.~III.B of the main text. For both materials, Li$_x$MnPO$_4$ and Li$_x$FePO$_4$, DFT fails to predict the correct change in the PDOS during the lithiation process due to SIE, and TM ions change their $d$ states occupation gradually and all together, assuming fractional oxidation states. Conversely, both DFT+$U$+$V$ and HSE06 show correct trends for these phospho-olivines. More specifically, when gradually adding Li ions to the materials, we see that only one TM ion at a time changes its OS from $+3$ to $+2$ which corresponds to a sudden jump in the (total) occupation of the corresponding Mn-$3d$ or Fe-$3d$ states. The overall similarity of the PDOS from DFT+$U$+$V$ and HSE06 is striking, even thought, as discussed in the main text, for Li$_x$Mn$_{1/2}$Fe$_{1/2}$PO$_4$ the fine details differ. In particular, the distribution of occupied states near the top of the valence bands is quite different within these two approaches: the intensity of the Mn-$3d$ states in the case of Li$_x$MnPO$_4$ and of Fe-$3d$ states in the case of Li$_x$FePO$_4$ is higher in HSE06, and there is a gap between them at $x=1$. 

It is useful to make a connection between the local geometry changes due to the pseudo Jahn-Teller distortions (see Sec.~\ref{sec:lattice_parameters}) and changes in the PDOS and atomic occupations. In fact, as can be seen in Fig.~\ref{fig:PDOS_LMPO} for Li$_x$MnPO$_4$ at $x=0$ there are empty states at around 2~eV above the top of the valence bands that are not present at $x=1$. It was shown in Ref.~\cite{Asari:2011SM} that these empty states originate from the $d_{xz}$ and $d_{x^2-y^2}$ states (these are the projections in the global Cartesian framework) in the Jahn-Teller active Mn$^{3+}$ ion. Since these $d_{xz}$ and $d_{x^2-y^2}$ states appear at the same energies, they are likely  hybridized~\cite{Asari:2011SM}. Our DFT+$U$+$V$ and HSE06 calculations confirm these findings. Hence, the emergence of empty states at 2~eV is the consequence of the pseudo Jahn-Teller distortions of the MnO$_6$ octahedra when delithiating Li$_x$MnPO$_4$ (thus going from the non-Jahn-Teller active Mn$^{2+}$ to the Jahn-Teller active Mn$^{3+}$). Moreover, as can be easily observed from Table~I of the main text, the smallest eigenvalue of the atomic occupation matrix in the spin-up channel ($\lambda_1^\uparrow$) equals to 0.50 in DFT+$U$+$V$, meaning that the corresponding eigenstate is half-empty which corresponds to the empty states appearing at about 2~eV. 

It is important to remark that our findings are consistent with previous computational studies for Li$_x$MnPO$_4$ ($x = 0, 1/4, 1/2, 3/4, 1$)~\cite{Piper:2013SM}. In particular, changes in the PDOS upon lithiation/delithiation are in good agreement between our DFT+$U$+$V$ and HSE06 predictions and those of DFT+$U$ investigations of Ref.~\cite{Piper:2013SM}. Following the thorough analysis of the structural, electronic, and spectroscopic properties of Li$_x$MnPO$_4$ using DFT+$U$ in combination with soft and hard X-ray spectroscopy~\cite{Piper:2013SM}, it would be important to perform similar measurements for Li$_x$FePO$_4$ and Li$_x$Mn$_{1/2}$Fe$_{1/2}$PO$_4$ that would help to verify the accuracy of the DFT+$U$+$V$ and HSE06 predictions presented here.

\newpage
\clearpage

\section{Band gaps}
\label{sec:band_gaps}

\begin{figure*}[h!]
  \includegraphics[width=0.8\linewidth]{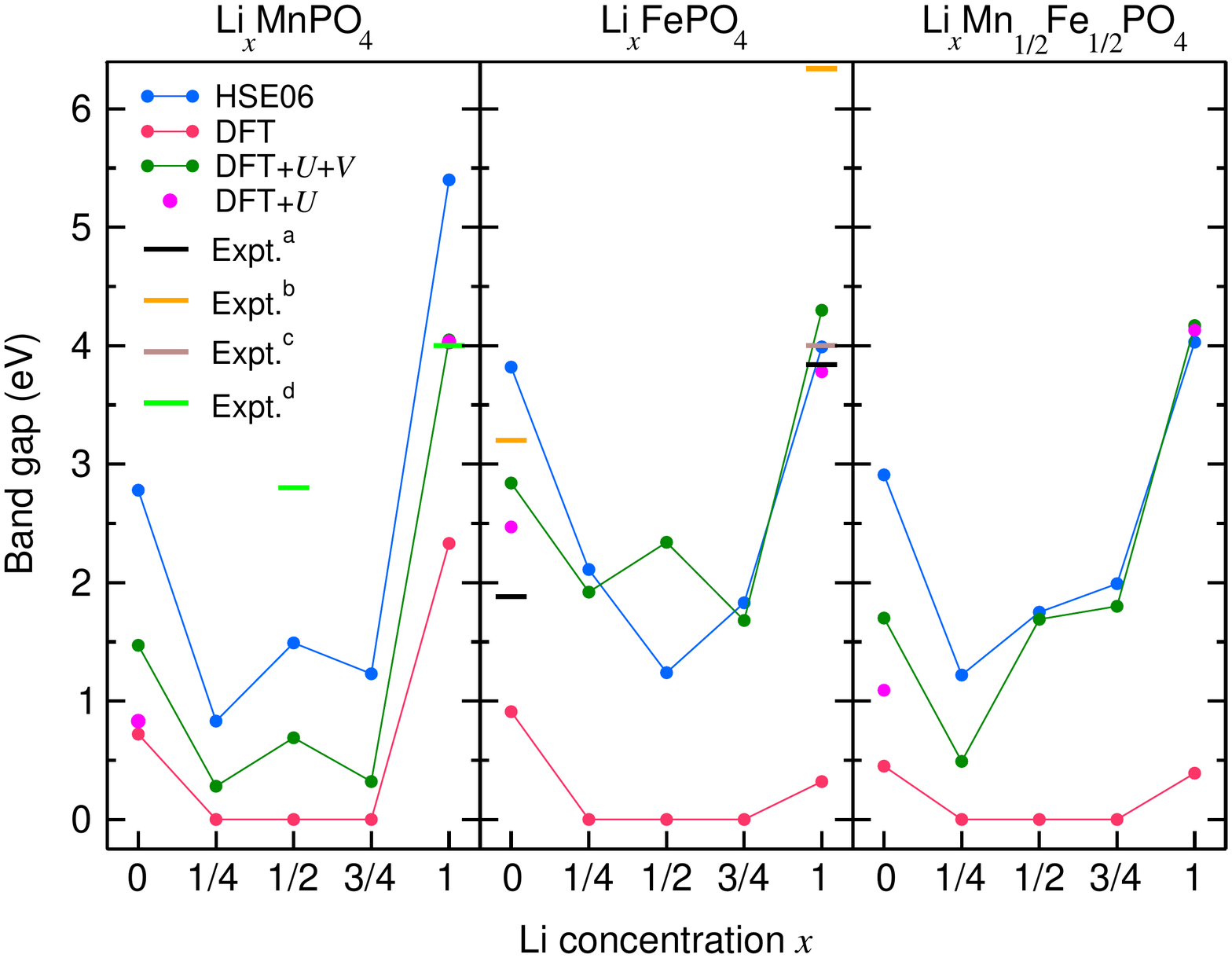}
   \caption{Band gaps for Li$_x$MnPO$_4$, Li$_x$FePO$_4$, and Li$_x$Mn$_{1/2}$Fe$_{1/2}$PO$_4$ computed using four approaches (DFT, DFT+$U$, DFT+$U$+$V$, and HSE06) and as measured in experiments: Expt.$^\mathrm{a}$ is Ref.~\onlinecite{Zaghib:2007SM}, Expt.$^\mathrm{b}$ is Ref.~\onlinecite{Zhang:2019SM}, Expt.$^\mathrm{c}$ is Ref.~\onlinecite{Zhou:2004cSM}, and Expt.$^\mathrm{d}$ is Ref.~\onlinecite{Piper:2013SM}.}
\label{fig:band_gap}
\end{figure*}

Figure~\ref{fig:band_gap} and Table~\ref{tab:band_gaps} show the band gaps of Li$_x$MnPO$_4$, Li$_x$FePO$_4$, and Li$_x$Mn$_{1/2}$Fe$_{1/2}$PO$_4$ in dependence of Li content, as computed from DFT, DFT+$U$, DFT+$U$+$V$, and HSE06 in comparison with experimental measurements from Refs.~\cite{Zaghib:2007SM, Zhang:2019SM, Zhou:2004cSM, Piper:2013SM}. It can be seen that the values of the band gaps are very sensitive to the method that is used. Unfortunately, experimental measurements of band gaps in these materials are relatively scarce, which makes it quite difficult to assess the accuracy of the considered approaches in predicting this quantity. Experimental values of the band gap were only found for Li$_x$MnPO$_4$ at $x=1/2$ and $x=1$~\cite{Piper:2013SM}, and for Li$_x$FePO$_4$ at $x=0$ and $x=1$~\cite{Zhou:2004cSM, Zaghib:2007SM, Zhang:2019SM}. Based on this data we can see that for Li$_x$FePO$_4$ at $x=0$ the DFT+$U$ and DFT+$U$+$V$ predictions fall in-between the two experimental values while the HSE06 band gap overestimates both of them. At $x=1$ we see that the DFT+$U$, DFT+$U$+$V$, and HSE06 band gaps are all close to the experimental values of Refs.~\cite{Zhou:2004cSM, Zaghib:2007SM}, while the experimental value of Ref.~\cite{Zhang:2019SM} is larger by $\sim 2$~eV. Instead, for Li$_x$MnPO$_4$ at $x=1$ there is a remarkable agreement between our DFT+$U$ and DFT+$U$+$V$ band gaps and the experimental value reported in Ref.~\cite{Piper:2013SM}, while HSE06 and DFT largely overestimate and underestimate the experimental gap, respectively. Curiously, for Li$_x$MnPO$_4$ at $x=1/2$, both DFT+$U$+$V$ and HSE06 significantly underestimate the experimental gap of Ref.~\cite{Piper:2013SM}. 

It is useful to compare general trends in the way band gaps change upon lithiation of these cathode materials. We observe that the DFT+$U$+$V$ and HSE06 band gaps behave quite similarly when changing $x$. However, in Li$_x$MnPO$_4$ we can see there is approximately a rigid downward shift of the DFT+$U$+$V$ band gaps compared to the HSE06 ones. Instead, in Li$_x$FePO$_4$ the DFT+$U$+$V$ and HSE06 band gaps are very close at  $x=1/4,3/4,1$ while at $x=0$ and $x=1/2$ they differ by $\sim 1$~eV. Conversely, in Li$_x$Mn$_{1/2}$Fe$_{1/2}$PO$_4$ the DFT+$U$+$V$ and HSE06 band gaps agree closely at $x=1/2,3/4,1$ while at $x=0$ and $x=1/4$ they differ by $\sim 1$~eV. The DFT+$U$ band gaps at $x=0$ and $x=1$ for different phospho-olivines are generally smaller than the DFT+$U$+$V$ ones, with a few exceptions when the two match (i.e. at $x=1$ for Li$_x$MnPO$_4$ and Li$_x$Mn$_{1/2}$Fe$_{1/2}$PO$_4$). Finally, the DFT predictions of band gaps differ notably from those of DFT+$U$+$V$ and HSE06: at fractional $x$ all materials are predicted to be metallic. This failure of DFT is due to the fact that Mn-$3d$ and Fe-$3d$ electrons are overdelocalized due to SIE and hence this leads to the closure of the gap.

A very instructive observation that emerges from the comparison of the results for Li$_x$Mn$_{1/2}$Fe$_{1/2}$PO$_4$ with those for Li$_x$MnPO$_4$ and Li$_x$FePO$_4$ is that for both DFT+$U$+$V$ and HSE06 the band gap (and its dependence on Li content) is indicative of the specific TM species that is correspondingly change its OS. In fact, for both these approaches, the band gap values for $0 \leq x < 1/2$ are very close to those of Li$_x$MnPO$_4$ (in the same range of Li content) while for $1/2 < x \leq 1$ they resemble those of Li$_x$FePO$_4$ for the same values of $x$. The only exception to this trend is represented by the DFT+$U$+$V$ band gap of Li$_x$Mn$_{1/2}$Fe$_{1/2}$PO$_4$ at $x=1/2$ whose value is intermediate between those of Li$_x$MnPO$_4$ and Li$_x$FePO$_4$ at the same $x$. This observation is fully consistent with (and further confirms) the fact that only one TM ion gets reduced at a time, and its $3d$ orbitals act as the frontier orbitals of the whole system and thus become responsible for its fundamental gap. This is true independently from their chemical environment, as if they were local states embedded in the bath represented by the rest of the crystal.

\begin{table*}[t]
\centering
\renewcommand{\arraystretch}{1.5}
\begin{tabular}{c|c|ccccc}
\hline\hline
Material & \parbox{1cm}{$x$}  & \parbox{2cm}{DFT} &  \parbox{1cm}{HSE06} & \parbox{2.5cm}{DFT+$U$+$V$} & \parbox{1cm}{DFT+$U$} & \parbox{3cm}{Expt.} \\ 
\hline
\parbox[t]{7mm}{\multirow{5}{*}{\rotatebox[origin=c]{90}{Li$_x$MnPO$_4$}}} 
&  0  & 0.72 & 2.78 & 1.47 & 0.83 &         \\
& 1/4 & 0.00 & 0.83 & 0.28 &      &         \\
& 1/2 & 0.00 & 1.49 & 0.69 &      & 2.8$^d$ \\
& 3/4 & 0.00 & 1.23 & 0.32 &      &         \\
&  1  & 2.33 & 5.40 & 4.05 & 4.03 & 4.0$^d$ \\ \hline
\parbox[t]{7mm}{\multirow{5}{*}{\rotatebox[origin=c]{90}{Li$_x$FePO$_4$}}} 
&  0  & 0.91 & 3.82 & 2.84 & 2.47 & 1.88$^a$, 3.2$^b$   \\
& 1/4 & 0.00 & 2.11 & 1.92 &      &     \\
& 1/2 & 0.00 & 1.24 & 2.34 &      &     \\
& 3/4 & 0.00 & 1.83 & 1.68 &      &     \\
&  1  & 0.32 & 3.99 & 4.30 & 3.78 & 3.84$^a$, 6.34$^b$, 4.00$^c$   \\ \hline
\parbox[t]{7mm}{\multirow{5}{*}{\rotatebox[origin=c]{90}{Li$_x$Mn$_{1/2}$Fe$_{1/2}$PO$_4$}}} 
&  0  & 0.45 & 2.91 & 1.70 & 1.09 &     \\
& 1/4 & 0.00 & 1.22 & 0.49 &      &     \\
& 1/2 & 0.00 & 1.75 & 1.69 &      &     \\
& 3/4 & 0.00 & 1.99 & 1.80 &      &     \\
&  1  & 0.39 & 4.03 & 4.17 & 4.13 &     \\
\hline\hline
\end{tabular}
\vskip 0.1cm
\hspace{-0.2cm} $^a$Ref.~\cite{Zaghib:2007SM}
\hspace{0.2cm}  $^b$Ref.~\cite{Zhang:2019SM}
\hspace{0.2cm}  $^c$Ref.~\cite{Zhou:2004cSM}
\hspace{0.2cm}  $^d$Ref.~\cite{Piper:2013SM}
\caption{Band gaps for Li$_x$MnPO$_4$, Li$_x$FePO$_4$, and Li$_x$Mn$_{1/2}$Fe$_{1/2}$PO$_4$ computed at different Li concentrations ($x=1,1/4,1/2,3/4,1$) using DFT, HSE06, DFT+$U$, and DFT+$U$+$V$, and as measured in experiments. We do not report the DFT+$U$ band gaps at fractional $x$ because of difficulties in stabilizing the $U$ values using the self-consistent protocol as explained in the main text (see Sec.~III.A).}
\label{tab:band_gaps}
\end{table*}


\begin{thebibliography}{182}
\bibitem [{Hub({\natexlab{a}})}]{Hubparam:2022}%
  \BibitemOpen
  \href@noop {} {} \bibinfo {note} {{The actual values
  produced by each of first-principles approaches for determining $U$ and $V$
  may vary substantially, in addition to the intrinsic dependence on the choice
  of the projector functions for the Hubbard manifold, pseudopotentials,
  oxidation state, functional, etc.}}\BibitemShut {Stop}%
\end{thebibliography}

%

\end{document}